\documentclass{an}
\usepackage{graphicx}
\usepackage{url}
\usepackage{times}
\newcommand{\EQ}{\begin{equation}}
\newcommand{\EN}{\end{equation}}
\newcommand{\EQA}{\begin{eqnarray}}
\newcommand{\ENA}{\end{eqnarray}}
\newcommand{\eq}[1]{(\ref{#1})}
\newcommand{\EEq}[1]{Equation~(\ref{#1})}
\newcommand{\Eq}[1]{Eq.~(\ref{#1})}
\newcommand{\Eqs}[2]{Eqs~(\ref{#1}) and~(\ref{#2})}

\newcommand{\Sec}[1]{Sect.~\ref{#1}}

\newcommand{\Fig}[1]{Fig.~\ref{#1}}
\newcommand{\FFig}[1]{Figure~\ref{#1}}
\newcommand{\Tab}[1]{Table~\ref{#1}}
\newcommand{\Figs}[2]{Figs~\ref{#1} and \ref{#2}}

\newcommand{\bra}[1]{\langle #1\rangle}
\newcommand{\bbra}[1]{\left\langle #1\right\rangle}

\newcommand{\meanemf}{\overline{\mbox{\boldmath ${\cal E}$}} {}}
\newcommand{\meanAA}{\overline{\bf{A}}}
\newcommand{\meanBB}{\overline{\bf{B}}}
\newcommand{\meanJJ}{\overline{\bf{J}}}
\newcommand{\meanUU}{\overline{\bf{U}}}
{}
{}
{}
{}
{}
{}
{}
\newcommand{\zz}{\hat{\mbox{\boldmath $z$}} {}}
\newcommand{\xx}{\mbox{\boldmath $x$} {}}

\newcommand{\zzz}{\mbox{\boldmath $z$} {}}

\newcommand{\bp}{\mbox{\boldmath $p$} {}}
\newcommand{\qq}{\mbox{\boldmath $q$} {}}

\newcommand{\uu}{{\bf{u}}}
\newcommand{\BB}{{\bf{B}}}
\newcommand{\JJ}{{\bf{J}}}

\newcommand{\AAA}{{\bf{A}}}

\newcommand{\bb}{{\bf{b}}}

\newcommand{\ee}{\mbox{\boldmath $e$} {}}

\newcommand{\hh}{\mbox{\boldmath $h$} {}}
\newcommand{\EE}{{\bf{E}}}

\newcommand{\kk}{\mbox{\boldmath $k$} {}}
\newcommand{\SSS}{{\bf{S}}}

\newcommand{\nab}{\mbox{\boldmath $\nabla$} {}}

\newcommand{\oo}{\mbox{\boldmath $\omega$} {}}

\newcommand{\ii}{{\rm i}}

\newcommand{\dd}{{\rm d} {}}

\def\la{\mathrel{\mathchoice {\vcenter{\offinterlineskip\halign{\hfil
$\displaystyle##$\hfil\cr<\cr\sim\cr}}}
{\vcenter{\offinterlineskip\halign{\hfil$\textstyle##$\hfil\cr<\cr\sim\cr}}}
{\vcenter{\offinterlineskip\halign{\hfil$\scriptstyle##$\hfil\cr<\cr\sim\cr}}}
{\vcenter{\offinterlineskip\halign{\hfil$\scriptscriptstyle##$\hfil\cr<\cr\sim\cr}}}}}

\newcommand{\ea}{{\rm et al.\ }}

\def\half{{\textstyle{1\over2}}}

\def\onethird{{\textstyle{1\over3}}}

\newcommand{\kG}{\,{\rm kG}}

\newcommand{\s}{\,{\rm s}}

\newcommand{\cm}{\,{\rm cm}}

\newcommand{\km}{\,{\rm km}}

\newcommand{\Mm}{\,{\rm Mm}}
\newcommand{\Mx}{\,{\rm Mx}}

\newcommand{\yr}{\,{\rm yr}}

\newcommand{\yjgr}[3]{ #1, {JGR,} {#2}, #3}
\newcommand{\ysol}[3]{ #1, {Sol. Phys.,} {#2}, #3}
\newcommand{\yapj}[3]{ #1, {ApJ,} {#2}, #3}
\newcommand{\yapjl}[3]{ #1, {ApJ,} {#2}, #3}
\newcommand{\yapjs}[3]{ #1, {ApJ Suppl.,} {#2}, #3}
\newcommand{\yan}[3]{ #1, {AN,} {#2}, #3}

\newcommand{\yana}[3]{ #1, {A\&A,} {#2}, #3}

\newcommand{\ygafd}[3]{ #1, {Geophys. Astrophys. Fluid Dyn.,} {#2}, #3}
\newcommand{\yjfm}[3]{ #1, {JFM,} {#2}, #3}

\newcommand{\yprs}[3]{ #1, {Proc. Roy. Soc. Lond.,} {#2}, #3}
\newcommand{\yprl}[3]{ #1, {PRL,} {#2}, #3}
\newcommand{\yphl}[3]{ #1, {Phys. Lett.,} {#2}, #3}

\newcommand{\ymn}[3]{ #1, {MNRAS,} {#2}, #3}
\newcommand{\ynat}[3]{ #1, {Nat,} {#2}, #3}

\newcommand{\ysph}[3]{ #1, {Solar Phys.,} {#2}, #3}
\newcommand{\ypr}[3]{ #1, {Phys. Rev.,} {#2}, #3}

\newcommand{\yjour}[4]{ #1, {#2}, {#3}, #4}

\newcommand{\ybook}[3]{ #1, {#2} (#3)}
\newcommand{\yproc}[5]{ #1, in {#3}, ed. #4 (#5), #2}

\newcommand{\pgafd}[1]{ #1, {Geophys. Astrophys. Fluid Dyn.,} (in press)}

\newcommand{\pjour}[2]{ #1, {#2,} (in press)}

\begin{document}
\lhead[\thepage]{A.~Brandenburg et al.: Magnetic helicity in stellar dynamos}
\rhead[Astron. Nachr./AN~{\bf XXX} (200X) X]{\thepage}
\headnote{Astron. Nachr./AN {\bf 32X} (200X) X, XXX--XXX}

\title{Magnetic helicity in stellar dynamos: new numerical experiments}
\author{Axel Brandenburg\inst{1}
  \and Wolfgang Dobler\inst{2}%
    \thanks{current address:
      Kiepenheuer Institute for Solar Physics, Sch\"oneckstr. 6, 79104
      Freiburg, Germany}
  \and Kandaswamy Subramanian\inst{3}%
    \thanks{current address:
      Inter University Centre for Astronomy and Astrophysics,
      Post Bag 4, Pune University Campus, Ganeshkhind, Pune 411 007, India}
}

\institute{
NORDITA, Blegdamsvej 17, DK-2100 Copenhagen \O, Denmark
\and
Department of Mathematics, University of Newcastle upon Tyne, NE1 7RU, UK
\and
National Centre for Radio Astrophysics - TIFR, Pune University Campus,
Ganeshkhind, Pune 411 007, India
}

\date{Received November 29, 2001;
%accepted {\it date will be inserted by the editor}}
revised \today,~ $ $Revision: 1.129 $ $}

\abstract{
The theory of large scale dynamos is reviewed with particular emphasis
on the magnetic helicity constraint in the presence of closed
and open boundaries. In the presence of closed or periodic boundaries,
helical dynamos respond to the helicity constraint by developing
small scale separation in the kinematic regime, and by showing
long time scales in the nonlinear regime where the scale separation
has grown to the maximum possible value. 
A resistively limited evolution towards saturation is also found at
intermediate scales before the largest scale of the system is reached.
Larger aspect ratios can give rise to different structures of
the mean field which are obtained at early times, but the final saturation
field strength is still decreasing with decreasing resistivity.
In the presence of shear, cyclic magnetic fields are found whose period
is increasing with decreasing resistivity, but the saturation energy of
the mean field is in strong super-equipartition with the turbulent energy.
It is shown that artificially induced losses of small scale field
of opposite sign of magnetic helicity as the large scale field
can, at least in principle, accelerate the production of large scale
(poloidal) field.
Based on mean field models with an outer potential field boundary
condition in spherical geometry, we verify that the sign of the magnetic
helicity flux from the large scale field
agrees with the sign of $\alpha$. For solar parameters,
typical magnetic helicity fluxes lie around $10^{47}\Mx^2$ per cycle.
}

\correspondence{brandenb@nordita.dk}

\maketitle

\section{Introduction}

The conversion of kinetic into magnetic energy, i.e.\ the dynamo effect,
plays an important role in many astrophysical bodies (stars, planets,
accretion discs, for example).
The generation of magnetic
fields on scales similar to the scale of the turbulence is a rather
generic phenomenon that occurs for sufficiently large magnetic Reynolds
numbers unless
certain antidynamo theorems apply, which exclude
for example two-dimensional fields (e.g., Cowling 1934).

There are two well-known mechanisms, which allow the generation of
magnetic fields on scales larger than the eddy scale of the turbulence:
the alpha-effect (Steenbeck,
Krause \& R\"adler 1966) and the inverse cascade of magnetic helicity in
hydromagnetic turbulence (Frisch \ea 1975,
Pouquet, Frisch, \& L\'eorat 1976). In a way the two
may be viewed as the same mechanism in that both are driven
by
helicity. The $\alpha$-effect is however clearly nonlocal
in wavenumber space,
whereas the inverse cascade of magnetic helicity is usually understood
as a (spectrally) local transport of magnetic energy to larger length scales.
Both aspects are seen in simulations of isotropic, non-mirror
symmetric turbulence: the nonlocal inverse
cascade or $\alpha$-effect in the early kinematic stage and the local
inverse cascade at later times, when the field has reached saturation at
small scales (Brandenburg 2001a, hereafter referred to as B01).

It is not clear whether either of these two mechanisms is actually
involved in the generation of large scale magnetic fields
in astrophysical bodies. Although much of the large scale magnetic field
in stars and discs
is the result of shear and differential rotation, a mechanism to sustain
a poloidal (cross-stream) magnetic field of sufficiently large scale is
still needed. The main problem that we shall be concerned with
in this paper is that of the associated magnetic helicity production,
which may prevent the large scale dynamo
process from operating on a dynamical time scale. However, not many
alternative and working dynamo mechanisms have been suggested so far.
In the solar context, the Babcock--Leighton mechanism is often discussed
as a mechanism that operates preferentially in the strong field regime
(e.g.\ Dikpati \& Charbonneau 1999).
However, Stix (1974) showed that the model of Leighton (1969)
is formally equivalent to the model of Steenbeck \& Krause
(1969), except that in Leighton (1969) the $\alpha$-effect was nonlinear
and mildly singular at the poles.
Nevertheless, a magnetically driven $\alpha$ effect has been suspected
to operate in turbulent flows driven by the magnetorotational instability
(Brandenburg \ea 1995) or by a magnetic buoyancy instability (Ferriz-Mas,
Schmitt, \& Sch\"ussler 1994, Brandenburg \& Schmitt 1998).
In the framework of mean-field dynamo theory, magnetically driven
dynamo effects have also been invoked to explain the observed increase
of stellar cycle frequency with increased stellar activity
(Brandenburg, Saar, \& Turpin 1998).
However, as long as these mechanisms lead to an $\alpha$-effect, they
also produce magnetic helicity and are hence subject to the same
problem as before.

Although kinetic helicity is crucial in the usual explanation of
the $\alpha$ effect, it is not a necessary requirement for magnetic field
generation, and lack of parity invariance
is already sufficient (Gilbert, Frisch, \& Pouquet 1988). However, as in every
$\alpha$ effect dynamo, magnetic helicity is produced at large scales,
which is necessarily a slow process. A mechanism similar to the
ordinary $\alpha$ effect is the {\it incoherent} $\alpha$ effect (Vishniac
\& Brandenburg 1997), which works in spite of constantly changing sign
of kinetic helicity (in space and time) provided there is systematic shear
and sufficient turbulent diffusion. This mechanism has been invoked
to explain the large scale magnetic field found in simulations of
accretion disc turbulence (Brandenburg \ea 1995, Hawley, Gammie, \&
Balbus 1996, Stone \ea 1996). In addition, however, the large scale field
of Brandenburg \ea (1995) shows spatio-temporal coherence with field
migration away from the midplane. This has so far only been possible
to explain in terms of an $\alpha\Omega$ dynamo with a magnetically
driven $\alpha$ effect (Brandenburg 1998). In any case, the incoherent
$\alpha$ effect, which has so far only been verified in one-dimensional
models, would not lead to the production of net magnetic helicity in
each hemisphere.

Yet another mechanism was suggested recently
by Vishniac \& Cho (2001),
which yields a mean electromotive force that does not lead to the
production of net magnetic helicity, but only to a transport of preexisting
magnetic helicity.
This is why this mechanism can, at least in principle,
work on a fast time scale even when the magnetic
Reynolds number is large. Finally, we mention a totally different
mechanism that works on the basis of negative turbulent diffusion (Zheligovsky,
Podvigina, \& Frisch 2001).
This is what produces turbulence in the Kuramoto--Sivashinsky Equation,
which is then stabilised by hyperdiffusion (i.e.\ a fourth derivative term).
Among all these different dynamo mechanisms,
the conventional $\alpha$ effect and the
inverse magnetic cascade are the only mechanisms that have been shown
numerically to produce strong large scale fields under turbulent conditions.
These are however exactly
the mechanisms that suffer from the helicity constraint.

\begin{figure}
\setlength{\fboxsep}{5pt}       % space between box and text
\setlength\unitlength{0.5\columnwidth}
\begin{center}
  \fbox{\begin{minipage}{0.6\columnwidth}
          \centering
          \vspace*{2ex}
          {\bf Large scale dynamos}\\
          (e.g.~in the sun)
          \vspace*{2ex}
        \end{minipage}}\\
  \addvspace{0.3ex}
  \begin{picture}(1,0.15)
    \put(0.5,0.15){\vector(0,-1){0.15}}
  \end{picture}\\
  \addvspace{0.3ex}
  \fbox{\begin{minipage}{0.6\columnwidth}
          \centering
          Is significant net magnetic helicity produced in each hemisphere?
        \end{minipage}}\\
  \addvspace{0.3ex}
  \begin{picture}(1,0.1)
    \put(0.1,0.1){\vector(-3,-2){0.18}}
    \put(0.5,0.1){\vector(-0,-2){0.10}}
    \put(0.90,0.1){\vector(3,-2){0.15}}
  \end{picture}\\
  \addvspace{0.3ex}
  \fbox{\begin{minipage}[l]{0.27\columnwidth}
          \raggedright
          \center{\it NO}\\
          fast $\alpha$ effect:\\
          (like in kinematic case,\\ and large scale separation)
        \end{minipage}}
  \hspace{\stretch{1}}
  \fbox{\begin{minipage}[l]{0.27\columnwidth}
          \raggedright
          \center{\it NO}\\
          non-$\alpha$ effect:\\
          (e.g., negative magn.\ diffusion,
          incoherent $\alpha$,
          Vishniac \& Cho effect)
        \end{minipage}}
  \hspace{\stretch{4}}
  \fbox{\begin{minipage}{0.27\columnwidth}
          \raggedright
          \center{\it YES}\\[2ex]
          standard $\alpha$ effect:\\
          $\rightarrow$ what happens to magnetic helicity?
        \end{minipage}}\\
  \addvspace{0.3ex}
  \begin{picture}(1,0.1)
    \put(-0.25,0.14){\vector(0,-1){0.15}}
    \put(+0.40,0.1){\vector(-3,-2){0.15}}
    \put(1.00,0.15){\vector(-1,-1){0.18}}
    \put(1.30,0.15){\vector(0,-2){0.18}}
  \end{picture}\\
  \addvspace{0.3ex}
  \fbox{\begin{minipage}[l]{0.18\columnwidth}
          \raggedright
          \center{Alternative~A}\\
          fast $\alpha$ effect: transfer from small to large scales,
          no net helicity
        \end{minipage}}
  \hspace{\stretch{1}}
  \fbox{\begin{minipage}[l]{0.18\columnwidth}
          \raggedright
          \center{Alternative~B}\\
          non-$\alpha$ effect dynamos: no numerical evidence yet
        \end{minipage}}
  \hspace{\stretch{4}}
  \fbox{\begin{minipage}{0.18\columnwidth}
          \center{Alternative~C}\\
          (i) dissipation\\
          (ii) reconnection\\
          (near surface?)
        \end{minipage}}
  \hspace{\stretch{1}}
  \fbox{\begin{minipage}{0.18\columnwidth}
          \center{Alternative~D}\\
          loss through boundaries:\\
          (i) equator?\\
          (ii) outer surface
        \end{minipage}}\\
\end{center}
\caption{Sketch illustrating the different ways astrophysical dynamos may
  be able to circumvent the magnetic helicity problem.
}\label{Fsketch}\end{figure}
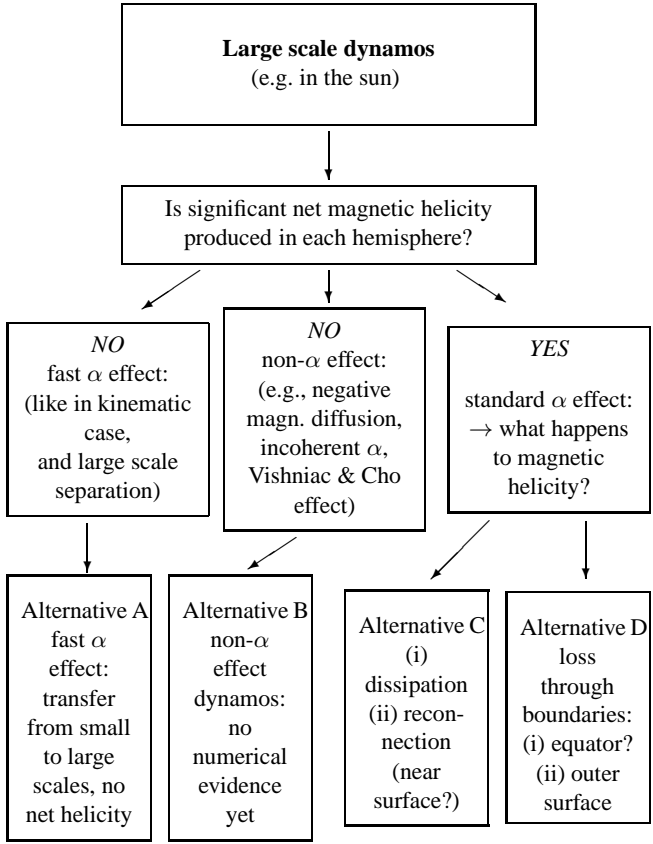

The purpose of the present paper is to assess the significance of the
helicity constraint and to present some new numerical experiments that
help understanding how the constraint operates and to discuss possible ways out of
the dilemma. For orientation and later reference we summarise in \Fig{Fsketch} various
models and the possible involvement of magnetic helicity in them.
Dynamos based on the usual $\alpha$ effect are listed to the
right. They all produce large scale magnetic fields that are helical.
However, in a periodic or an infinite domain, or in the presence of
perfectly conducting boundaries, the magnetic helicity is conserved and
can only change through microscopic resistivity. This would therefore seem to be
too slow for explaining variations of the mean field on the time scale
of the 11 year solar cycle. Open boundary conditions may help, although
this was not yet possible to demonstrate, as will be explained in
\Sec{Shelevol} of this paper.

%% The diagram

Given that the issue of magnetic helicity is central to everything
that follows, we give in \Sec{Srecap} a brief review of magnetic helicity
conservation, the connection between the inverse cascade and the
realisability condition, and the significance of
gauge-invariant forms of magnetic helicity and the surface integrated
magnetic helicity flux. Readers familiar with this may jump directly
to \Sec{Shelproblem}, where we discuss the issue of magnetic helicity
cancellation in kinematic and non-kinematic dynamos, or
to \Sec{Shelevol}, where we begin with a discussion of the results of
B01 and Brandenburg \& Dobler (2001, hereafter referred to as BD01),
or to \Sec{Sremoval} where new results are presented.

\section{The magnetic helicity constraint}
\label{Srecap}

Magnetic helicity evolution does not give any additional information 
beyond that already contained in the induction equation
governing the evolution
of the magnetic field itself. Nevertheless, the concept of magnetic
helicity proves extremely useful because magnetic helicity is a conserved 
quantity in ideal MHD, and almost conserved even in slightly non-ideal
conditions. The evolution equation for the magnetic
helicity can then be used to extract information that can be easily understood.
At a first glance, however, magnetic helicity appears counterintuitive,
because it involves the magnetic vector potential, $\AAA$, which is not
itself a physical quantity, as it is not invariant under the gauge
transformation $\AAA\rightarrow\AAA-\nab\psi$. Only the magnetic field,
$\BB=\nab\times\AAA$, which lacks the irrotational `information' from
$\AAA$, is independent of $\psi$ and hence physically meaningful.
Therefore $\AAA\cdot\BB$ is not a physically meaningful quantity either,
because a gauge transformation, $\AAA\rightarrow\AAA-\nab\psi$ changes
$\AAA\cdot\BB$. For periodic or perfectly conducting boundary conditions,
or in infinite domains, the integral $H=\int\AAA\cdot\BB\,\dd V$ is however
gauge-invariant, because in
\EQ
\int_V\nab\psi\cdot\BB\,\dd V
=\int_{\partial V}\psi\BB\cdot\dd\SSS
-\int_V\psi\nab\cdot\BB\,\dd V
\label{psi-B}
\EN
the surface integral on the right hand side vanishes, and the last term
vanishes as well, because $\nab\cdot\BB = 0$.

The evolution of the magnetic field is governed by the induction equation
\EQ
{\partial\BB\over\partial t}=-\nab\times\EE,
\label{dBdt}
\EN
where $\EE$ is the electric field. This equation can be integrated to yield
\EQ
{\partial\AAA\over\partial t}=-\EE-\nab\Phi,
\label{dAdt}
\EN
where $\Phi$ is the scalar (`electrostatic') potential,
which is also referred to
as the gauge potential, because it can be chosen arbitrarily without
affecting $\BB$.
Common choices are $\nabla^2\Phi=-\nab\cdot\EE$, which preserves the
Coulomb gauge $\nab\cdot\AAA=0$, and $\Phi=0$, which is convenient in MHD for
numerical purposes. If $\eta=\mbox{const}$, another convenient gauge is
$\Phi=-\eta\nab\cdot\AAA$, which results in
\EQ
  \frac{\partial\AAA}{\partial t}
  = \uu\times\BB + \eta\nabla^2\AAA
\quad\mbox{($\eta=\mbox{const}$)}.
 \label{Induction-AAA}
\EN
Using \Eqs{dBdt}{dAdt} one can derive an equation for the gauge-dependent
magnetic helicity density $\AAA\cdot\BB$,
\EQ
{\partial\over\partial t}(\AAA\cdot\BB)=-2\EE\cdot\BB
-\nab\cdot(\EE\times\AAA+\Phi\BB).
\label{dABdt}
\EN
Using Ohm's law, $\eta\mu_0\JJ = \EE + \uu\times\BB$,
we have $2\EE\cdot\BB=2\eta\mu_0\JJ\cdot\BB$, where $\eta$
is the microscopic magnetic diffusivity, $\mu_0$ the magnetic permeability
in vacuum, and $\JJ=\nab\times\BB/\mu_0$ the electric current density.

\subsection{Magnetic helicity conservation with periodic boundaries}

Consider first the case of a periodic domain, so the divergence term in
\Eq{dABdt} vanishes after integration over the full volume, and thus
\EQ
{\dd\over\dd t}\bra{\AAA\cdot\BB}=
-2\eta\mu_0\bra{\JJ\cdot\BB},
\label{helcons}
\EN
where angular brackets denote volume averages, so
$\int\AAA\cdot\BB\,\dd V=\bra{\AAA\cdot\BB}V$,
where $V$ is the (constant) volume of the domain under consideration.
Note that
the terms in \Eq{helcons} are gauge-independent [see \Eq{psi-B}].
More important, however,
is the fact that the rate of change of magnetic helicity is
proportional to the \emph{microscopic} magnetic diffusivity $\eta$.
This alone is not sufficient to conclude that the magnetic
helicity will not change in the limit $\eta\rightarrow0$,
because the current helicity, $\bra{\JJ\cdot\BB}$, may still
become large. A similar effect is encountered in the case
of ohmic dissipation of magnetic energy which proceeds at the
rate $Q_{\rm Joule}\equiv\eta\mu_0\bra{\JJ^2}$. In the statistically steady state
(or on time averaging), the rate of ohmic dissipation
must be balanced by the work done against the Lorentz force.
Assuming that this work term is independent of the value of $\eta$
(e.g.\ Galsgaard \& Nordlund 1996)
we conclude that the root-mean-square current density increases with
decreasing $\eta$ like
\EQ
J_{\rm rms}\propto\eta^{-1/2}\quad
\mbox{as $\eta\rightarrow0$},
\EN
whilst the rms magnetic field strength, $B_{\rm rms}$, is essentially
independent of $\eta$.
This, however, implies that the rate of magnetic helicity
dissipation {\it decreases} with $\eta$ like
\EQ
Q_H\equiv2\eta\mu_0\bra{\JJ\cdot\BB}V\propto\eta^{+1/2}\rightarrow0
\quad\mbox{as $\eta\rightarrow0$}.
\label{DHchange}
\EN
Thus, under many astrophysical conditions where the magnetic Reynolds number is large
($\eta$ small), the magnetic helicity $H$, as governed by \Eq{helcons},
is almost independent of time. This motivates the search for dynamo
mechanisms independent of $H$. On the other hand,
it is conceivable that magnetic helicity of one sign can change
on a dynamical time scale, i.e.\ faster than in \Eq{DHchange}, if
magnetic helicity of the other sign is removed locally by advection,
for example. This is essentially what the mechanism of Vishniac \& Cho
(2001) is based upon. So far, however, numerical attempts to
demonstrate the operation of this mechanism have failed
(Arlt \& Brandenburg 2001).
Yet another possibility is that magnetic helicity of one sign 
is generated at large scales, and is compensated by magnetic helicity
of the other sign at small scales, as in the kinematic  $\alpha$ effect.
However, as we shall see in \Sec{Shelevol}, this does not happen efficiently when nonlinear
effects of the Lorentz force come into play.

\subsection{Realisability condition and connection with inverse cascade}
\label{Srealisability}

Magnetic helicity is important for large scale field generation because
it is a conserved quantity which cannot easily cascade forward to smaller
scale. We shall demonstrate this here for the case where the magnetic field
is fully helical. The existence of an upper bound for the magnetic
helicity is easily seen by decomposing the Fourier transformed magnetic vector
potential, $\AAA_{\kk}$, into a longitudinal component, $\hh^\parallel$, and
eigenfunctions $\hh^{\pm}$ of the curl operator (also called
Chandrasekhar--Kendall functions),
\EQ
\AAA_{\kk}=a_{\kk}^+\hh_{\kk}^++a_{\kk}^-\hh_{\kk}^-
+a_{\kk}^\parallel\hh_{\kk}^\parallel,
\EN
with
\EQ
  \nabla\times\hh_{\kk}^{\pm} = \pm k \hh_{\kk}^{\pm},
  \quad\quad k = |\kk|,
\EN
and
\EQ
\bra{{\hh_{\kk}^+}^*\cdot\hh_{\kk}^+}
=\bra{{\hh_{\kk}^-}^*\cdot\hh_{\kk}^-}
=\bra{{\hh_{\kk}^{\parallel}}\cdot{\hh_{\kk}^{\parallel}}}=1,
\EN
where asterisks denote the complex conjugate.
The longitudinal part $a_{\kk}^\parallel\hh_{\kk}^\parallel$ is parallel to $\kk$ and
vanishes after taking
the curl to calculate the magnetic field.
In the Coulomb gauge, $\nabla\cdot\AAA=0$, the longitudinal component vanishes
altogether (apart from an uninteresting constant vector $\AAA_0$).
Since $\hh^{\pm}$ and $\hh^\parallel$ are all orthogonal to each
other, it is clear that, for a given magnetic field,
the Coulomb gauge minimises $\bra{\AAA^2}$
--- or, if $\AAA_0=\bra{\AAA}\ne0$, the variance
$\bra{(\AAA{-}\bra{\AAA})^2}$.

The (complex) coefficients
$a_{\kk}^\pm(t)$ depend on $\kk$ and $t$, while the eigenfunctions
$\hh_{\kk}^\pm$, which form an orthonormal set, depend only on $\kk$
and are given by\footnote{
  The forcing functions used
  in B01 were proportional to $\hh_{\kk}^+$.
  We note that, by mistake, the square root
  was missing in the normalisation factor $1/\sqrt{2}$ in that paper.}
\EQ
\hh_{\kk}^\pm=\frac{1}{\sqrt{2}}{\kk\times(\kk\times\ee)\mp\ii
k(\kk\times\ee)\over k^2\sqrt{1-(\kk\cdot\ee)^2/k^2}},
\EN
where $\ee$ is an arbitrary unit vector not parallel to $\kk$.
With this definition
we can write the magnetic helicity and energy spectra in the form
\begin{eqnarray}
  H_k &=& k(|a^+|^2-|a^-|^2)V,
  \label{H-apm}\\
  M_k &=& \half k^2(|a^+|^2+|a^-|^2)V ,
  \label{M-apm}
\end{eqnarray}
where $V$ is the volume of integration.
(Here and elsewhere the factor $\mu_0^{-1}$ is ignored in the definition
of the magnetic energy.) These spectra are normalised such that
\EQ
\int_0^\infty H_k\,\dd k=\bra{\AAA\cdot\BB}V\equiv H,
\EN
\EQ
\int_0^\infty M_k\,\dd k=\bra{\half\BB^2}V\equiv M,
\EN
where $H$ and $M$ are magnetic helicity and magnetic energy,
respectively. From \Eqs{H-apm}{M-apm} one sees immediately that
\EQ
\half k|H_k|\leq M_k,
\label{realisability}
\EN
which is also known as the {\it realisability condition}.
A fully helical field has therefore $M_k=\pm\half k H_k$.

For the following it is convenient to write the
magnetic helicity spectrum in the form
\EQ
H_k=\bra{\AAA_k\cdot\BB_k}V/\delta k,
\label{helspectrum}
\EN
where the subscript $k$ (which is here a scalar!)
indicates Fourier filtering to only those
wave vectors $\kk$ that lie in the shell
\EQ
k-\delta k/2\ge|\kk|<k+\delta k/2\quad(\mbox{$k$-shell}).
\EN
Note that $\AAA_k$ and $\BB_k$ are in real (configuration) space,
so $\BB_k=\nab\times\AAA_k$.
(By contrast, $\AAA_{\kk}$ and $\BB_{\kk}=\ii\kk\times\AAA_{\kk}$ are quantities in
Fourier space, as indicated by the bold face subscript $\kk$, which is
here a vector.)
In \Eq{helspectrum} the average is taken over all points in $\xx$ space.
Note that in practice the helicity is more easily calculated directly as
\EQ
H_k=\frac{1}{2}\int_{\mbox{$k$-shell}}(\AAA_{\kk}^*\cdot\BB_{\kk}+
\AAA_{\kk}\cdot\BB_{\kk}^*)\,\kk^2\,\dd\Omega_{\kk},
\label{helspectrum_fft}
\EN
where $\dd\Omega_{\kk}$ is the solid angle element in Fourier space.
The result is however the same.
Likewise, the magnetic energy spectrum can
be written as
\EQ
M_k=\half\bra{\BB_k^2}V/\delta k,
\EN
or as
\EQ
M_k = \frac{1}{2}\int_{\mbox{$k$-shell}}
      \BB_{\kk}^*\cdot\BB_{\kk}\,k^2\,d\Omega_k.
\EN
We recall that for a periodic domain $H$ is gauge invariant. Since
its spectrum can be written as an integral over all space, see
\Eq{helspectrum}, $H_k$ is -- like $H$ -- also gauge invariant.

The occurrence of an inverse cascade can be understood as the result of
two waves (wavenumbers
$\bp$ and $\qq$) interacting with each other to produce a wave of
wavenumber $\kk$. The following argument is due to Frisch \ea (1975).
Assuming that during this process magnetic energy is
conserved together with magnetic helicity, we have
\EQ
M_p + M_q = M_k ,
\label{frisch_energy}
\EN
\EQ
|H_p| + |H_q| = |H_k| ,
\label{helconspqk}
\EN
where we are assuming that only helicity of one sign is involved.
Suppose the initial field is fully helical and has the same
sign of magnetic helicity at all scales, then we have
\EQ
2M_p=p|H_p|\quad\mbox{and}\quad
2M_q=q|H_q|,
\EN
and so \Eq{frisch_energy} yields
\EQ
p |H_p| + q|H_q| = 2M_k \ge k|H_k| ,
\EN
where the last inequality is just the realisability condition
\eq{realisability} applied to
the target wavenumber $\kk$ after the interaction. Using \Eq{helconspqk}
we have
\EQ
p|H_p|+q|H_q|\ge k(|H_p|+|H_q|).
\EN
In other words, the target wave vector $\kk$ after the interaction of
wavenumbers $\bp$ and $\qq$ satisfies
\EQ
k\le{p|H_p|+q|H_q|\over|H_p|+|H_q|}.
\label{helconspqk2}
\EN
The expression on the right hand side of \Eq{helconspqk2} is a weighted
mean of $p$ and $q$ and thus satisfies
\EQ
\min(p,q)\le{p|H_p|+q|H_q|\over|H_p|+|H_q|}\le\max(p,q),
\EN
and therefore
\EQ
k\le\max(p,q).
\EN
In the special case where $p=q$, we have $k\le p=q$, so the target
wavenumber after interaction is always less or equal to the initial
wavenumbers. In other words, wave interactions tend to transfer magnetic
energy to smaller wavenumbers, i.e.\ to larger scale. This corresponds
to an inverse cascade. The realisability condition, $\half k|H_k|\le M_k$,
was perhaps the most important assumption in this argument.
Another important assumption that we
made in the beginning was that the initial field be fully helical;
see \Sec{Sfractional} for the case of fractional helicity.

We note that in hydrodynamics, without magnetic fields, the kinetic
helicity spectrum, $F_k=\bra{\oo_k\cdot\uu_k}V/\delta k$, where
$\oo=\nab\times\uu$ is the vorticity, is also conserved
by the nonlinear terms, but now the realisability condition is
\EQ
\half k^{-1}|F_k|\le E_k,
\EN
so it is not $k$, but the inverse of $k$, that enters the inequality. Here,
$E_k=\half\bra{\uu_k^2}V$ is related to the kinetic energy spectrum
(i.e.\ without density), and hence
one has
\EQ
k^{-1}\le{p^{-1}|F_p|+q^{-1}|F_q|\over|F_p|+|F_q|},
\EN
or
\EQ
k^{-1}\le\max(p^{-1},q^{-1}).
\EN
Thus $k \ge \min(p,q)$, and
in the special case $p=q$ we have $k\ge p=q$, so kinetic energy
is transferred preferentially to larger wavenumbers, so there is no
inverse cascade in this case.

\subsection{Spectra of right and left handed components}

For further reference we now define power spectra of those components
of the field that are either right or left handed, i.e.\
\EQ
H_k^\pm=\pm k|a_\pm|^2V,\quad
M_k^\pm=\half k^2|a_\pm|^2V.
\EN
Thus, we have $H_k=H_k^++H_k^-$ and $M_k=M_k^++M_k^-$. Note that $H_k^\pm$
and $M_k^\pm$ can be calculated without explicit decomposition into
right and left handed field components using
\EQ
H_k^\pm=\half(H_k\pm2k^{-1}M_k),\quad
M_k^\pm=\half(M_k\pm\half kH_k).
\EN
This method is significantly simpler than invoking explicitly the
decomposition in terms of $a_{\kk}^\pm\hh_{\kk}^\pm$.

\subsection{Open boundaries and the importance of a gauge-independent magnetic helicity}
\label{Sopenbc}

Let us now discuss the case where magnetic helicity
is allowed to flow through boundaries.
The first problem we encounter in connection with open boundaries is
that the magnetic helicity is no longer gauge invariant. There are
however known procedures that allow $H$ to be defined in a way that
is independent of gauge. Before we
discuss this in more detail we first want to demonstrate how the use
of a gauge-dependent magnetic helicity can lead to undesired artifacts.

For illustrative purposes we consider the case of a one-dimensional
mean-field $\alpha^2$ dynamo governed by the equation
\EQ
{\partial\meanAA\over\partial t}=\alpha\meanBB-\eta_{\rm T}\mu_0\meanJJ,
\label{dmeanAAdt}
\EN
where $\eta_{\rm T}=\eta_{\rm t}+\eta$ is the sum
of turbulent and microscopic magnetic diffusivity,
and $\meanBB=(\overline{B}_x,\overline{B}_y,0)=\meanBB(z)$ is the
magnetic field which depends only on $z$, and
$\meanBB=\nab\times\meanAA$ as well as $\meanJJ=\nab\times\meanBB/\mu_0$.
Here we have used the gauge $\Phi=0$.
The overbars denote horizontal
$(x,y)$ averages and the corresponding mean quantities like $\overline{B}$
are therefore independent of $x$ and $y$.
As boundary conditions we adopt either
$\partial\overline{A}_x/\partial z=\partial\overline{A}_y/\partial z=0$
(vacuum boundary condition) or, for comparison,
$\overline{A}_x=\overline{A}_y=0$ (perfect conductor condition)
on $z=0,L_z$. We adopt $\alpha$ and $\eta_{\rm t}$ quenching, i.e.\
\EQ
\alpha={\alpha_0\over1+a\meanBB^2/B_{\rm eq}^2},\quad
\eta_{\rm t}={\eta_{\rm t0}\over1+a\meanBB^2/B_{\rm eq}^2}.
\label{quenching}
\EN
(In the case considered here we use $a=1$.)

%\begin{figure}[t!]\centering\includegraphics[width=0.5\textwidth]{fig/psum.ps}\caption{
\begin{figure}[t!]\centering\includegraphics[width=0.5\textwidth]{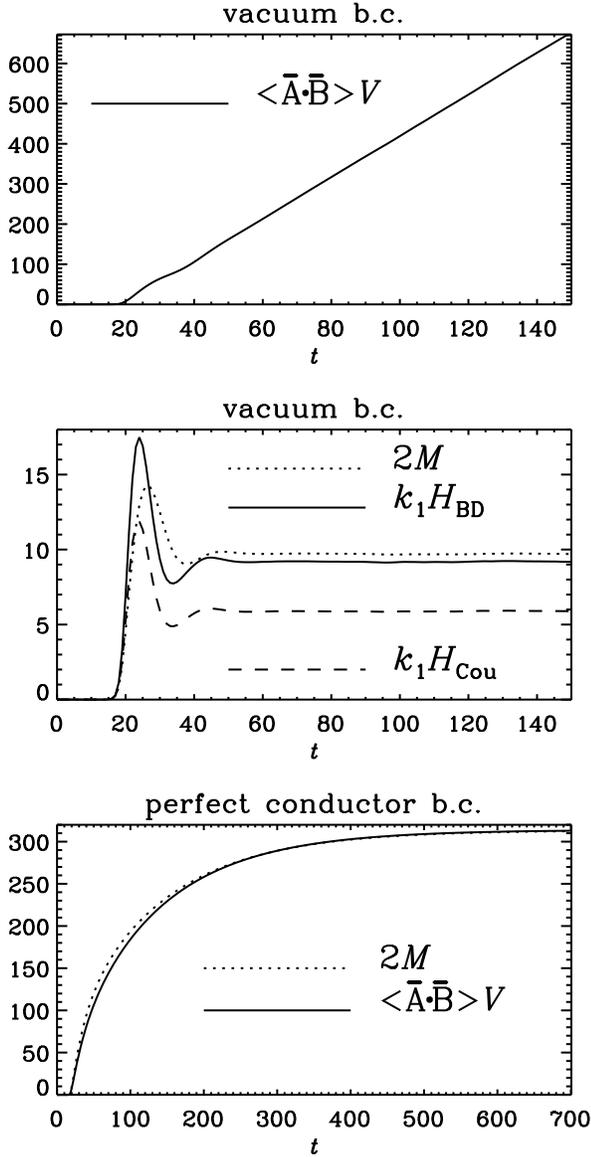}\caption{
First panel: gauge-dependent magnetic helicity for an $\alpha^2$ dynamo
with vacuum boundary conditions. Second panel: gauge-independent
magnetic helicities $H_{\rm BD}$ (solid line) and $H_{\rm Cou}$ (dashed line),
compared with twice the magnetic energy, $2M$, for the same model (dotted line).
Third panel: same as second panel, but with perfect conductor boundary
conditions. Here gauge-independent magnetic helicities agree with the
ordinary expression, $\int\meanAA\cdot\meanBB\,\dd V$.
Again, the dotted line shows $2M$.
}\label{Fpsum}\end{figure}

The result of a numerical integration of \Eq{dmeanAAdt} is shown in \Fig{Fpsum}.
Evidently, the use of a gauge-dependent magnetic helicity implies
spurious contributions that prevent $\bra{\meanAA\cdot\meanBB}$
from becoming time-independent
even though the magnetic energy has long reached a steady state.
The reason for this is that for the steady state solution the right hand
side of \Eq{dmeanAAdt} is a constant which is different from zero.
As an example, consider the eigenfunction of the linearised equations:
$\meanBB=(1+\cos k_1z,\sin k_1z,0)B_0$, where $k_1=1$ for $L_z=2\pi$.
In the marginally excited state, we have $\alpha=\eta_{\rm T}k_1$,
so $\partial{\overline A}_x/\partial t=\alpha{\overline B}_0(t)\neq0$.
Because of this complication,
it is important to define magnetic helicity in a
gauge-independent way.
BD01 defined such a gauge-independent magnetic helicity by
linearly extending the horizontally averaged magnetic vector potential to
a periodic field which then results in a gauge-invariant magnetic helicity
as before; see \Eq{psi-B}.
This can be written in compact form as
\EQ
H_{\rm BD}=\bra{\meanAA\cdot\meanBB}V+\zz\cdot(\meanAA_1\times\meanAA_2)V,
\label{HBD}
\EN
where $\meanAA_1$ and $\meanAA_2$ are the values of $\meanAA$ on
the lower and upper boundaries, and angular brackets denote full volume
averages, as opposed to the overbars, which denote only horizontal
averages.\footnote{We use this opportunity
to point out a sign error in Eq.~(9) of BD01, where it should read
$\AAA_0=\overline{\AAA}_0 -\nab_\perp\times(\psi\zz)$. The contributions
from the mean field, that are central both here and in BD01,
remain however unaffected.}

We note that $H_{\rm BD}$ can be written in
the form proposed by Finn \& Antonsen (1985),
\EQ
H_{\rm BD}(\meanBB)=
\bra{(\meanAA+\meanAA_{\rm P})
\cdot(\meanBB-\meanBB_{\rm P})}V
\label{FA85form}
\EN
where $\meanBB_{\rm P}$ is the non-helical reference field
\EQ
\meanBB_{\rm P}=\bra{\BB},
\EN
which is spatially constant.
A corresponding $\meanAA_{\rm P}$ that satisfies
$\meanBB_{\rm P}=\nab\times\meanAA_{\rm P}$ is
\EQ
\meanAA_{\rm P}=-\zzz\times\meanBB_{\rm P}+\mbox{const},
\label{meanAAP}
\EN
which, apart from some constant, corresponds to a linear interpolation of $\meanAA$
between $z_1$ and $z_2$. However, the constant in \Eq{meanAAP} drops out
in the term $\meanAA_1\times\meanAA_2$ of \Eq{HBD}.
It is also easy to see that the constant drops out in \Eq{FA85form},
because $\bra{\meanBB-\meanBB_{\rm P}}=0$.
It is precisely this fact that makes \eq{FA85form} gauge-invariant.

In the following we compare with the magnetic helicity in the gauge
$\meanAA\mapsto\meanAA-\bra{\meanAA}$, which minimises $\bra{\meanAA^2}$,
\EQ
H_{\rm Cou}=\bra{\meanAA\cdot\meanBB}V-\bra{\meanAA}\cdot\bra{\meanBB}V.
\label{H-Cou}
\EN
This magnetic helicity is actually also gauge-invariant, but it cannot be written
in the form \eq{FA85form}, because it is not based on a reference field.
Indeed, the only value of $\meanBB_{\rm P}$ that makes the expression
\eq{FA85form} gauge-invariant is $\meanBB_{\rm P}=\bra{\meanBB}$,
but this does already corresponds to $H_{\rm BD}$.

We associate $H_{\rm Cou}$ with the Coulomb gauge because it is based on
a gauge which minimises $\bra{\AAA^2}$, which is characteristic of the
Coulomb gauge; see \Sec{Srealisability}.
However, in the present geometry, for horizontally averaged fields,
and zero net magnetic flux, a canonical
gauge transform $\meanAA\mapsto\meanAA-\zz\partial\psi/\partial z$ does not change
the large scale magnetic helicity $\bra{\meanAA\cdot\meanBB}$, since
$\overline{B}_z=0$.
In this sense, $\bra{\meanAA\cdot\meanBB}V$ in any arbitrary gauge
represents the magnetic helicity in Coulomb gauge.
There is, however, the extra freedom to add a constant vector $-\AAA_0$
to the magnetic vector potential, $\meanAA \mapsto \meanAA - \AAA_0$.
It is easy to see that $\bra{\meanAA^2}$ is minimised if
$\AAA_0 = \bra{\meanAA}$, and that the definition (\ref{H-Cou}) always
yields the magnetic helicity which corresponds to this gauge.

In the second
panel of \Fig{Fpsum} we have plotted both $H_{\rm BD}$ as well as
$H_{\rm Cou}$. Note that $|H_{\rm Cou}|\leq|H_{\rm BD}|$ and also
$|H_{\rm Cou}|\leq2M/k_{\min}$. If the domain was periodic, 
and since the computational domain is $-\pi<x<\pi$, we would have
$k_{\min}=1$. However, for vacuum boundary conditions one
could still accommodate a wave with $k_{\rm min}=1/2$. Then also $|H_{\rm BD}|$
is always less than $2M/k_{\min}$.

In the more general case of three-dimensional fields we adopt the
gauge-independent magnetic helicity of Berger \& Field (1984).
Their definition, however, leaves the contribution from the
average of $\AAA$ undetermined and can therefore only be applied
to the deviations from the mean field.

\subsection{Resistively limited growth and open boundaries}

A plausible interpretation of the
resistively limited growth is the following (Blackman \& Field 2000,
Ji \& Prager 2001). To satisfy magnetic helicity
conservation, the growth of the large scale field has to proceed
with very little (or very slow) changes of net magnetic helicity.
This is essentially the result of B01 (see also Brandenburg \&
Subramanian 2000, hereafter BS00). The underlying
assumption was that magnetic helicity of the sign opposite to that
of the mean field is lost resistively. As a working hypothesis
we assume that such losses occur preferentially at small
scales where also Ohmic losses are largest.

In order to speed up this process it might help to get rid of
magnetic helicity with sign opposite to that of the large scale
field. One possibility would be via explicit losses of field
with opposite magnetic helicity through boundaries (i.e.\ the
outer surface or the equatorial plane).
This possibility was first discussed by Blackman \& Field (2000)
and more recently by Kleeorin et al.\ (2000) and Ji \& Prager (2001).

The results obtained so far seem to indicate that this does not
happen easily (BD01). Instead, since most of the magnetic helicity
is already in large scales, most of the losses
are in large scales, too. Thus, the idea of losses at small scales,
where the sign of magnetic helicity is opposite to that at large
scale, has obvious difficulties. In \Sec{mhelsun} we address
the question of observational evidence for this. In \Sec{Shelevol}
we discuss quantitatively the effect of losses on large scales.

\section{Magnetic helicity cancellation}
\label{Shelproblem}

\subsection{Magnetic helicity in ABC-flow dynamos}
\label{SABC-flow}

\begin{figure}[t!]\centering\includegraphics[width=0.5\textwidth]{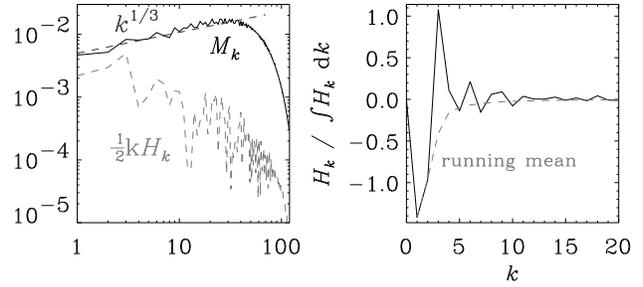}\caption{
Power spectrum, $M_k$, of the magnetic field for an {ABC} flow dynamo
with $A=B=C\equiv U_0=1$ and $k=1$. Note that most of the power is in the small
scales with $M_k\sim k^{1/3}$. The normalised helicity spectrum is
also plotted. Note that for medium and large values of $k$ the spectral
magnetic helicity is very small; $\half k|H_k|\ll M_k$.
The simulation was done with $240^3$ mesh points and a magnetic
diffusivity $\eta=2\times10^{-4}$.
}\label{Fppspe_bb}\end{figure}

\begin{figure}[t!]\centering\includegraphics[width=0.5\textwidth]{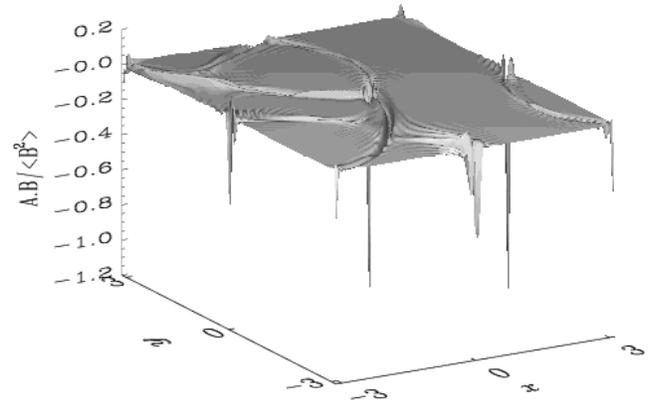}\caption{
Cross-section of the magnetic helicity density for the ABC flow dynamo
from \Fig{Fppspe_bb}. Note the intermittent nature of $\AAA\cdot\BB$.
}\label{Fpabsurf}\end{figure}

\begin{figure}[t!]\centering\includegraphics[width=0.5\textwidth]{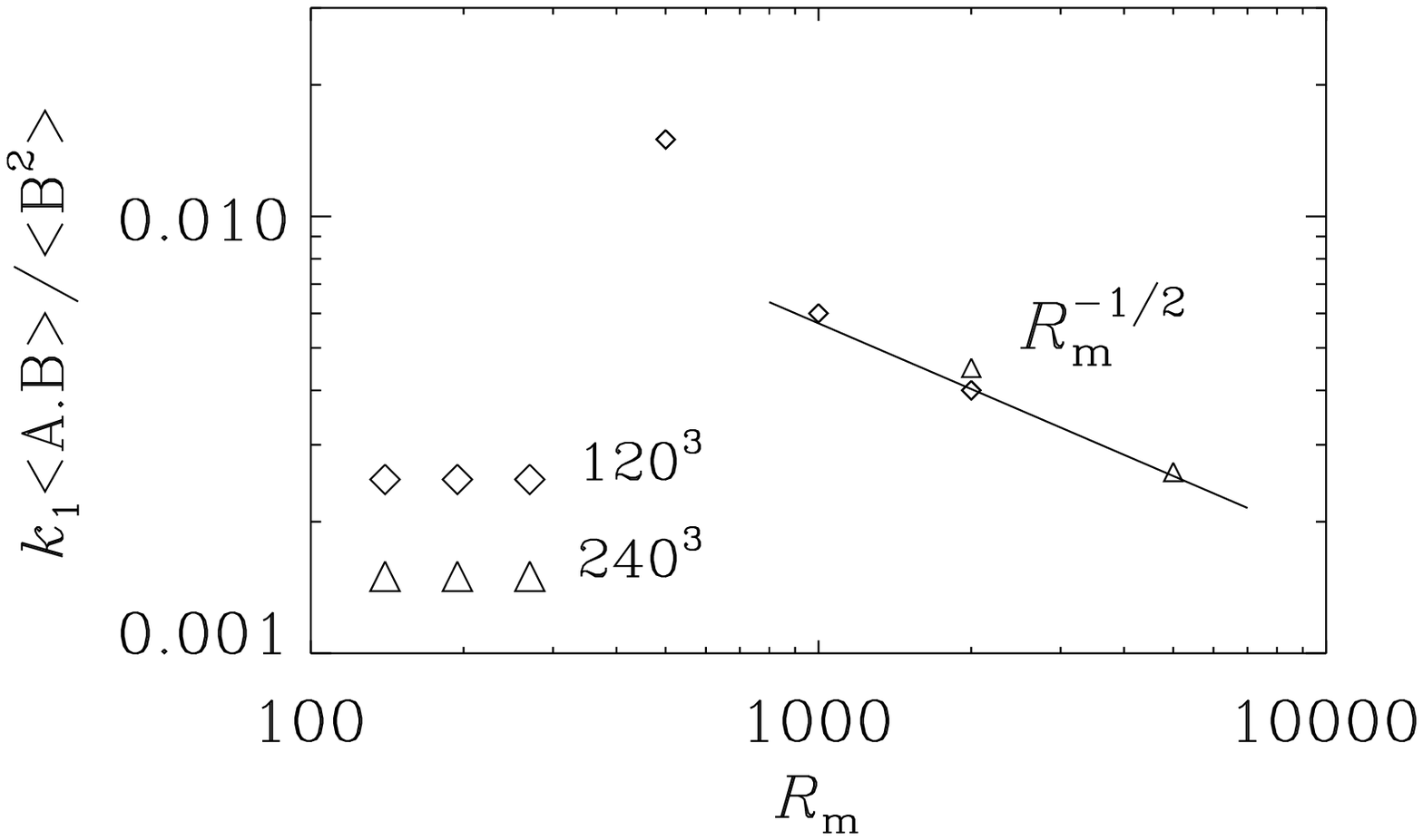}\caption{
Magnetic helicity, $\bra{\AAA\cdot\BB}$, nondimensionalised by $\bra{\BB^2}/k_1$,
as a function of $R_{\rm m}=U_0/(\eta k_1)$ for a model with $A=B=C\equiv U_0=1$.
Here, $k_1=1$ is the smallest wavenumber in the box, which
has the size $(2\pi)^3$. Note that $k_1\bra{\AAA\cdot\BB}/\bra{\BB^2}$ decreases
with increasing $R_{\rm m}$ like $R_{\rm m}^{-1/2}$. Thus, in the high magnetic Reynolds
number limit $1/\eta\to\infty$, the magnetic helicity of the {ABC}
flow dynamo vanishes.
}\label{Fpabc_rm}\end{figure}

Before we start discussing the evolution of magnetic helicity in turbulent
dynamos we consider first the case of steady {\it kinematic} ABC flows, where
the velocity field is given by
\EQ
\uu=\pmatrix{
A\sin k_1z+C\cos k_1y\cr
B\sin k_1x+A\cos k_1z\cr
C\sin k_1y+B\cos k_1x}
\EN
with $A=B=C=1$ and $k_1=1$ in the domain $-\pi<x,y,z<\pi$.
This class of dynamo models
is of interest because here the net magnetic helicity can be small,
even though the flow itself is fully helical (Gilbert 2002, and references therein).
The field generated by such an ABC flow dynamo does have
magnetic helicity, but it is comparatively weak in the sense that the
realisability condition \eq{realisability}
is far from being saturated.
This is seen in \Fig{Fppspe_bb}, where we compare $M_k$ with
$\half k|H_k|$.
In addition, the spectral helicity is alternating in sign, often from
one wavenumber to the next; see the second panel of \Fig{Fppspe_bb}. The
magnetic helicity density is also alternating in {\it space}, with a
smoothly varying, mostly positive component, interspersed by negative,
highly localised peaks as shown in \Fig{Fpabsurf}.
Moreover, as $R_{\rm m}\rightarrow\infty$,
the net magnetic helicity goes asymptotically to zero;
see \Fig{Fpabc_rm}. This result is typical of fast dynamos
(Hughes, Cattaneo, \& Kim 1996, Gilbert 2002).

%\begin{figure}[t!]\centering\includegraphics[width=0.5\textwidth]{fig/ppspe_bb.ps}\caption{

%\begin{figure}[t!]\centering\includegraphics[width=0.5\textwidth]{fig/pabsurf.ps}\caption{

%\begin{figure}[t!]\centering\includegraphics[width=0.5\textwidth]{fig/pabc_rm.ps}\caption{

We note in passing that the magnetic power spectrum increases approximately like $k^{1/3}$
(see the left panel of \Fig{Fppspe_bb}),
which is similar to the results from a number of other dynamos as long as they are in the
kinematic regime; see, e.g., Fig.~22 of B01 for a large magnetic Prandtl
number
calculation and Fig.~17 of Brandenburg \ea (1996) for convective dynamos.
Thus, most of the spectral magnetic energy is in small scales, which is
quite different from the helical turbulence simulations of B01.
In turbulent dynamos this spectrum could be explained by the analogy between
magnetic field and vorticity (Batchelor 1950). For a Kolmogorov $k^{-5/3}$
spectrum, the vorticity spectrum has a $k^{1/3}$ inertial range. In the present
case, however, this argument does not apply because the flow is stationary, so
there is no cascade in velocity. Thus, the origin of the $k^{1/3}$ spectrum
for these types of kinematic dynamos remains unclear.

\begin{figure}[t!]\centering\includegraphics[width=0.5\textwidth]{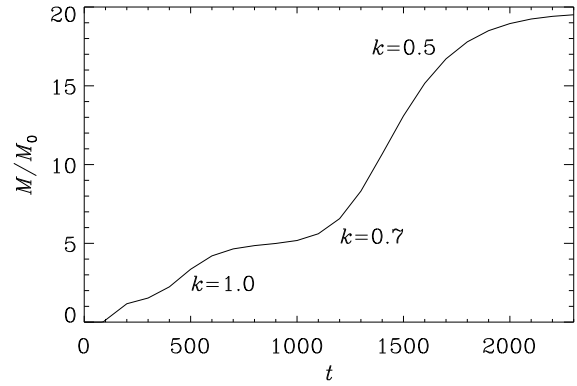}\caption{
Evolution of the magnetic energy $M=\bra{\BB^2/2\mu_0}$, for a `nonlinearised' {ABC} flow
with $A=0$, and $B=C=1/(1+M/M_0)$ scales with the
magnetic energy, $M$.
The box has a side length of $8\pi$ in the $z$ direction.
In the kinematic stage ($t < 500$) modes with $k>1$
grow fastest.
After the small scale modes saturate, the energy of the larger scales
continues to grow until the large scale reaches
saturation (e.g., $k=0.5$ saturates after $t=2000$).
The field continues to grow even further, but very slowly.
Since the box has a side length of $8\pi$, the minimum wavenumber is $k=0.25$.
}\label{Fpmm}\end{figure}

One might wonder whether nearly perfect magnetic helicity cancellation
can also occur in the sun. Before we can address this question we need
to understand the circumstances under which this type of magnetic helicity
cancellation can occur. It appears that nearly perfect magnetic helicity cancellation is
an artefact of the assumption of linearity and of
a small computational domain. Increasing its size allows
larger scale fields to grow even though the field has already saturated on the
scale of the {ABC} flow. This is shown in \Fig{Fpmm} where we plot the
magnetic energy of the mean magnetic field, $\meanBB$, for a `nonlinearised' {ABC}
flow with $A=0$, and $B=C=1/(1+M/M_0)$, where $M$ is the magnetic energy and $M_0=1$.
We chose a domain of size $2\pi\times2\pi\times8\pi$, which is increased in the
$z$ direction by a factor 4. In the kinematic phase the high wavenumber
modes grow fastest.  Once the magnetic energy reaches saturation, the
velocity is decreased, so the high wavenumber modes stop growing,
but the remaining lower wavenumber modes are still excited. Although
they grew slower during the earlier kinematic stage, they are now still excited and
saturate at a much later time. This process continues until the mode
with the geometrically largest possible scale has reached saturation,
provided that mode remains still excited.
Once a large scale field of this form has developed, cancellation of
any form has become small. This is indicated by the fact, that the
magnetic helicity is nearly maximum, i.e.\ $H/(2\mu_0M)\approx k_m^{-1}$
with $k_m$ being the {\it momentary} wavenumber of the large scale
magnetic field, which is also indicated in \Fig{Fpmm}.
Thus, for this type of dynamo, small scale magnetic fields are highly
intermittent and can grow fast, while large-scale fields grow once again
on slow time scales only.
It is this type of large scale field generation that is often thought to be
responsible for generating the large scale field of the sun (see
Parker 1979, Krause \& R\"adler 1980, or Zeldovich et al.\ 1983,
for conventional models of the
solar dynamo), although the saturation process is still resistively slow.

%\begin{figure}[t!]\centering\includegraphics[width=0.5\textwidth]{fig/pmm.ps}\caption{

\subsection{Magnetic helicity cancellation in turbulent dynamos}
\label{Subsec-cancellation}
\begin{figure}[t!]\centering\includegraphics[width=0.5\textwidth]{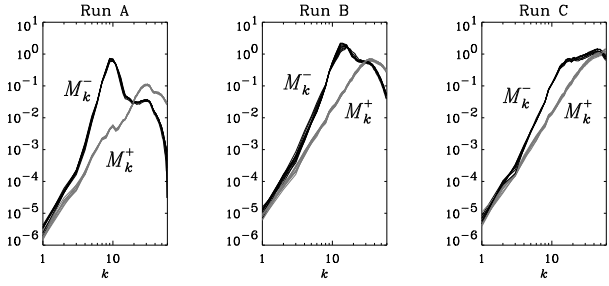}\caption{
Power spectra separately for the right and left handed parts
of the field, for Runs~A, B, and C of BS02. Different times
have been collapsed onto the same graph by scaling the spectra with
$e^{-\gamma t}$, where $\gamma$ is the corresponding growth rate
of the magnetic energy for each run. Note that as the magnetic
diffusivity is reduced, the relative excess of $M_k^-$ over $M_k^+$
decreases, as expected for asymptotic magnetic helicity conservation.
}\label{Fpspec_pm_kin_Hyp2b}\end{figure}

Magnetic helicity cancellation seems to be a generic phenomenon of all
{\it kinematic} dynamos, including the turbulent dynamos in cartesian domains
that have been studied recently in B01. Even though the initial field is
fully helical, the signs of magnetic helicity are different at large and
small scales and can therefore cancel, at least in principle. In practice
there remains always some finite net magnetic helicity, but as the magnetic
Reynolds number increases, the total magnetic helicity decreases, exactly
like for the ABC-flow dynamos studied in \Sec{SABC-flow}. Examples of magnetic power
spectra calculated separately for the right and left handed parts of the
field are shown in \Fig{Fpspec_pm_kin_Hyp2b} for Runs~A, B, and C of
Brandenburg \& Sarson (2002, hereafter referred to as BS02).
These are models with helical forcing around a wavenumber $k_{\rm f}$, like in B01.
The basic parameters of the runs presented in BS02 are summarised in \Tab{Tsum}.
Run~A has ordinary magnetic diffusivity with $\eta=10^{-4}$, whilst Runs~B and C
have second order hyperdiffusivity, so $\eta\nabla^2\AAA$ is replaced by
$-\eta_2\nabla^4\AAA$ in the induction equation (\ref{Induction-AAA}),
and $\eta_2=3\times10^{-8}$ and $10^{-8}$ for Runs~B and C, respectively.
[Generally, for models with $n$th order hyperdiffusivity, one would have
a diffusion term $(-1)^{n+1}\eta_n\nabla^{2n}\AAA$.]
In all three cases the
forcing of the flow is at wavenumber $k_{\rm f}=27$ with the same amplitude
($f_0=0.1$), as in B01. At the forcing scale, Run~B is by a factor
$\eta_1k_{\rm f}^2/(\eta_2 k_{\rm f}^4)=0.22$
times less resistive than Run~A. Run~C is another three times less
resistive.
During the kinematic phase shown, the spectral power increases at all wavenumbers
at a rate proportional to $e^{\gamma t}$, where $\gamma=2\lambda$
is the kinematic growth rate of the magnetic energy and $\lambda$ the growth
rate of the rms magnetic field.

%\begin{figure}[t!]\centering\includegraphics[width=0.5\textwidth]{fig/pspec_pm_kin_Hyp2b.ps}\caption{

\begin{figure}[t!]\centering\includegraphics[width=0.5\textwidth]{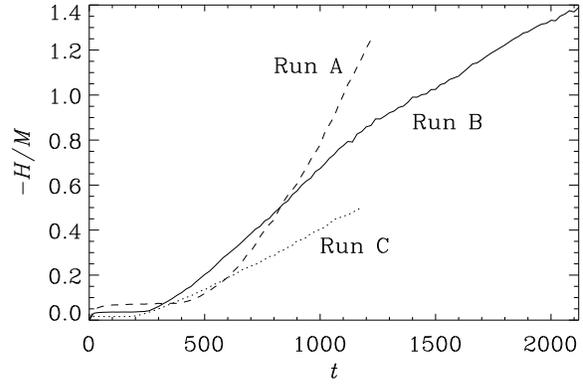}\caption{
Magnetic helicity, normalised by the magnetic energy, for Runs~A, B, and
C of BS02. In Run~A, $\eta=10^{-4}$, whereas in Runs~B and C second order
hyperresistivity is used, with $\eta_2=3\times10^{-8}$ and $10^{-8}$, respectively.
Note that the smaller the magnetic diffusivity (at large and intermediate scales),
the slower is the saturation of magnetic helicity.
}\label{Fppm_comp2}\end{figure}

As the magnetic Reynolds number increases, the magnetic helicity
normalised by the magnetic energy, $-H/M$, decreases, as shown in
\Fig{Fppm_comp2} for Runs~A, B, and C of BS02.
The helicity of the forcing and the kinetic helicity of the flow are
positive, so the magnetic helicity at small scales is also positive.
This results in large scale magnetic helicity that is negative.
Therefore, also the total magnetic helicity is negative.

The asymptotic decrease of magnetic helicity for progressively smaller
magnetic diffusivity (\Fig{Fpabc_rm}) can be quantitatively described by Berger's (1984)
magnetic helicity constraint, which we modify here for the initial
kinematic evolution when the magnetic energy increases exponentially
at the rate $\gamma=2\lambda$. We consider first the case of ordinary
magnetic diffusivity and turn then to the case of hyperdiffusivity.
For a closed or periodic domain, the modulus of the time derivative
of the magnetic helicity can be bounded from above by
\EQA
\begin{array}{lll}
\displaystyle{\frac{1}{\mu_0}
\left|{\dd H\over\dd t}\right|}=2\eta|\bra{\JJ\cdot\BB}|V
&\leq&2\eta\bra{\JJ^2}^{1/2}\bra{\BB^2}^{1/2}V\\
&=&2\eta^{1/2}Q_{\rm Joule}^{1/2}(2M)^{1/2},
\end{array}
\label{Bergers_relation}
\ENA
where $Q_{\rm Joule}=\bra{\eta\mu_0\JJ^2}$ is the rate of Joule dissipation.
In the kinematic regime, $|\dot{H}|=\gamma|H|$ and 
$\dot{M}=\gamma M=W-Q_{\rm Joule}=\epsilon Q_{\rm Joule}$, where $W$ is
the work done on the magnetic field by the velocity field.
In the dynamo case, i.e.\ if $\gamma>0$, we have $\epsilon>0$.
Thus
\EQ
{\gamma\over\mu_0}|H|\leq
2\eta^{1/2}(\gamma M/\epsilon)^{1/2}(2M)^{1/2}
=(2\eta\gamma/\epsilon)^{1/2}(2M)
\EN
or
\EQ
|H|/(2\mu_0 M)\leq \epsilon^{-1/2} (2\eta/\gamma)^{1/2}
=a\ell_{\rm skin},
\label{H-M-lskin}
\EN
where $\ell_{\rm skin}=(2\eta/\gamma)^{1/2}$
is the skin depth associated with the growth rate $\gamma$
and $a=\epsilon^{-1/2}$ is a parameter describing the
fractional excess of magnetic energy gain over dissipation.

In the case of hyperdiffusivity, which is often used for numerical
reasons, we have
\EQ
Q_{\rm Joule}\equiv\eta_n\int |\kk|^{2n}{|B_{\kk}|^2\over\mu_0}\,d^3k
\approx\eta_n k_{\rm f,eff}^{2n}\bbra{{\BB^2\over\mu_0}}V,
\label{approxdef_for_keff}
\EN
where $k_{\rm f,eff}\approx k_{\rm f}$ is an effective wavenumber defined
such that the second equality in \Eq{realdef_for_keff} below is satisfied
exactly and in \Eq{approxdef_for_keff} only approximately.
The rate of change of magnetic helicity is then bounded by
\EQA
\begin{array}{lll}
\displaystyle{{\gamma\over\mu_0}}|H|&\leq&
2\eta_n\int |\kk|^{2n-1}|B_{\kk}|^2/\mu_0\,d^3k\\
&\equiv&2\eta_n^{1/2} k_{\rm f,eff}^{n-1}
\bra{\eta_nk_{\rm f,eff}^{2n}\BB^2/\mu_0}^{1/2}\bra{\BB^2/\mu_0}^{1/2}V,
\end{array}
\label{realdef_for_keff}
\ENA
or
\EQA
\begin{array}{lll}
\displaystyle{{\gamma\over\mu_0}}|H|&\leq&
2(\eta_n k_{\rm f,eff}^{2n-2})^{1/2} Q_{\rm Joule}^{1/2}(2M)^{1/2}\\
&=&(2\eta_n k_{\rm f,eff}^{2n-2})^{1/2} (\gamma/\epsilon)^{1/2}(2M).
\end{array}
\ENA
So, again,
\EQ
|H|/(2M)\leq a\ell_{\rm skin},
\EN
but now $\ell_{\rm skin}=(2\eta_n k_{\rm f,eff}^{2n-2}/\gamma)^{1/2}$
involves the value of $k_{\rm f,eff}$. Like in the case with ordinary
magnetic diffusivity we have $a=\epsilon^{-1/2}$. From the values
of $|H|/(2\mu_0M)$ and $\ell_{\rm skin}$ given in \Tab{Tsum} we
see that this constraint is indeed well satisfied during the kinematic
growth phase.

\begin{table*}[htb]\caption{
Summary of runs, some of which were already presented in BS02.
The runs with $n=2$ have hyperdiffusivity,
so the effective diffusivity at the forcing wavenumber, $k_{\rm f}$, and at the
Nyquist wavenumber, $k_{\rm Ny}=\pi/\delta x=N/2$, are different.
$N^3$ is the total number of mesh points, and $\delta x$ is the mesh spacing.
The parameter $\ell_{\rm skin}$
gives an approximate upper bound for $|H|/(2\mu_0M)$.
}\begin{tabular}{lccccccccccc}
Run &   $N$   &$n$& $\eta_n$    &  $\eta_n k_{\rm f}^{2n-2}$ &  $\eta_n k_{\rm Ny}^{2n-2}$ & $\gamma$ & $k_{\rm f}$ & $\alpha_{\rm trav}$
&  $|H|/(2\mu_0M)$ & $\ell_{\rm skin}$ & $S/(2M)|_{\rm t=1000}$ \\
\hline
A&$120^3$&1&  $10^{-4}$     &  $10^{-4}$     &  $10^{-4}$     &0.047 &27&$1.1\times10^{-3}$& 0.035& 0.065&0.00075\\%Hyp2b A
B&$120^3$&2&$3\times10^{-8}$&$2\times10^{-5}$&  $10^{-4}$     & 0.070&27&$7.3\times10^{-4}$& 0.018& 0.025&0.00035\\%Hyp3b C
C&$120^3$&2&    $10^{-8}$   &$7\times10^{-6}$&$4\times10^{-5}$& 0.082&27&$3.6\times10^{-4}$& 0.005& 0.013&0.00020\\%Hyp3  D or C'
D& $30^3$&2&  $10^{-4}$     &$9\times10^{-4}$&$2\times10^{-2}$&0.078 & 3&      --          & 0.08 & 0.15 &  --  \\%hyp5d B
E&$ 60^3$&2&  $10^{-5}$     &$2\times10^{-4}$&$9\times10^{-3}$& 0.12 & 5&      --          & 0.040& 0.065&  --  \\%hhyp2f
F&$120^3$&2&$3\times10^{-8}$&$3\times10^{-7}$&  $10^{-4}$     & 0.12 & 3&      --          &0.0013&0.0021&  --  \\%Hyp5
\label{Tsum}\end{tabular}\end{table*}

%\begin{figure}[t!]\centering\includegraphics[width=0.5\textwidth]{fig/ppm_comp2.ps}\caption{

\subsection{Magnetic helicity conversion to large scales}

\begin{figure}[t!]\centering\includegraphics[width=0.5\textwidth]{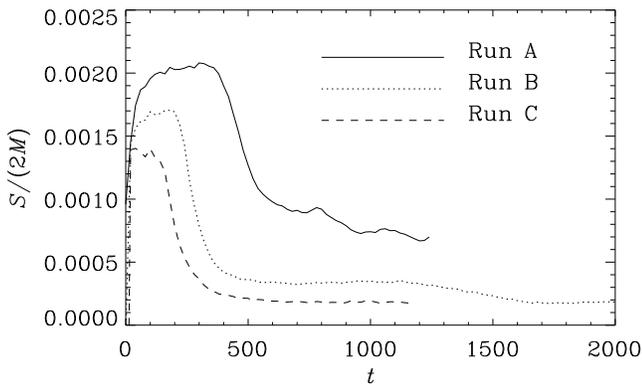}\caption{
Relative magnetic helicity transfer from small to large scales for runs A
(most resistive),
B and C (least resistive) of BS02, as a function of time.
Note that for smaller magnetic diffusivity, the ratio $S/(2M)$ decreases
and hence the conversion to large scales becomes less efficient.
}\label{Fppm_trans}\end{figure}

At least initially, i.e.\ when the field is still growing exponentially,
there is a strong conversion of positive magnetic
helicity at the driving scale to negative magnetic helicity at a certain larger scale.
This can be described by the magnetic helicity equation, written separately
for positive and negative parts,
\EQ
\dot{H}_\pm=\pm2S-2\eta k_\pm^2 H_\pm,
\label{Htransfer}
\EN
where $H_\pm=\int H_k^\pm\dd k$,
with $k_\pm^n=\int k^n H_k^\pm\dd k/\int H_k^\pm\dd k$, and
$S$ describes the transfer of magnetic helicity from small to large
scales. 
In runs with second order hyperresistivity, the term $\eta k_\pm^2$
has to be replaced by $\eta_2 k_\pm^4$.
The evolution of the net magnetic helicity, $H^++H^-$, is
of course unaffected by $S$. We note, however, that the {\it spectrum}
of the magnetic helicity does have non-diffusive source and sink terms
such that its integral over all wavenumbers vanishes (see B01 for
detailed plots). The term $\pm S$ in \Eq{Htransfer} is related
to the $\alpha$ effect (Blackman \& Field 2000) or, more precisely,
to the turbulent electromotive force. In the present case,
where turbulent diffusion also enters, we have therefore (cf.\ B01)
\EQ
S=\bra{\alpha\meanBB^2-\eta_{\rm t}\mu_0\meanJJ\cdot\meanBB}V.
\label{Sexpression}
\EN
Since the other two terms in \Eq{Htransfer} are known,
we can directly work out $S$ from this equation.
The results for Runs~A, B, and C are plotted in \Fig{Fppm_trans}.
The quantity $S/(2M)$ is a measure of the rate of helicity transfer,
i.e.\ the difference between $\alpha$-effect and turbulent diffusion.
Although the ratio $S/(2M)$ has not yet settled to a stationary value,
we compare in \Tab{Tsum} the values for three different runs (at $t=1000$).

%\begin{figure}[t!]\centering\includegraphics[width=0.5\textwidth]{fig/ppm_trans.ps}\caption{

\FFig{Fppm_trans} shows that $S/(2M)$ approaches a finite value in the nonlinear
regime, but its value becomes smaller as the magnetic diffusivity is decreased.
Thus, even though strong final field strengths are possible, this can still
be achieved with a residual $\alpha$ effect that decreases with decreasing
magnetic diffusivity. This is consistent with earlier results of
Cattaneo \& Vainshtein (1991), Vainshtein \& Cattaneo (1992), and
Cattaneo \& Hughes (1996) that $\alpha$ and $\eta_{\rm t}$ are quenched
in a $R_{\rm m}$-dependent fashion. Field \& Blackman (2002) have pointed
out, however, that this result can also be explained by a dynamically
quenched $\alpha$ where $\alpha$ and $\eta_{\rm t}$ are not necessarily
quenched in a $R_{\rm m}$-dependent fashion.

%\begin{figure}[t!]\centering\includegraphics[width=0.5\textwidth]{fig/pspec_allo_Hyp3b.ps}\caption{
\begin{figure}[t!]\centering\includegraphics[width=0.5\textwidth]{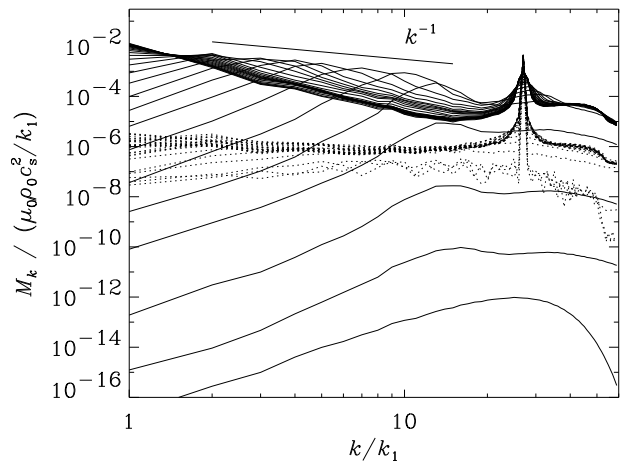}\caption{
Evolution of magnetic energy spectra in the hyperdiffusive Run~B
in equidistant time intervals
from $t=0$ (lowest curve), $t=40$ (next one higher up), until $t=2080$
(peaking at the very top left). The dotted lines give the kinetic energy.
}\label{Fpspec_allo_Hyp3b}\end{figure}

\begin{figure}[t!]\centering\includegraphics[width=0.5\textwidth]{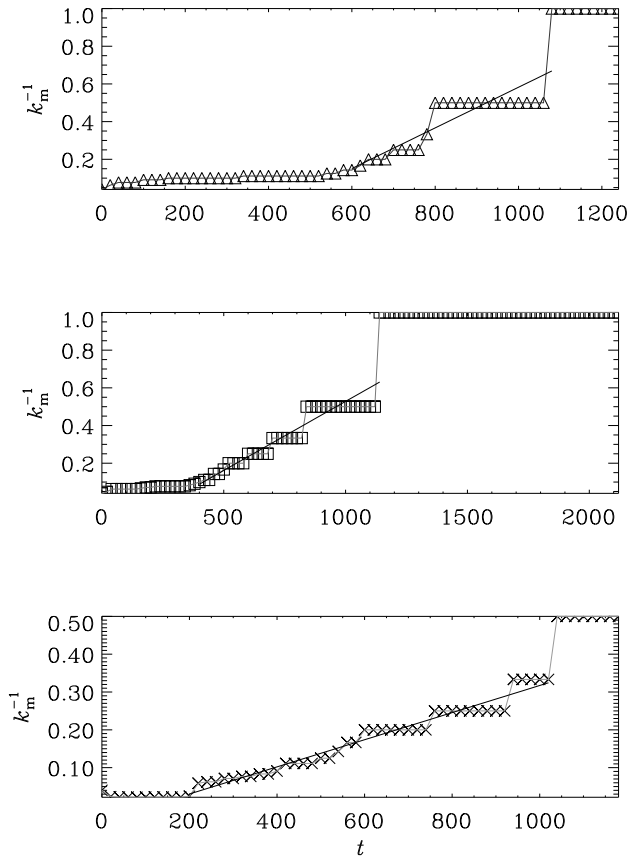}\caption{
Evolution of the position of the secondary bump in the power spectrum.
}\label{Fpkpeak_comp}\end{figure}

One may wonder whether the $R_{\rm m}$-dependent slow-down during the
saturation phase of the large scale field (as seen in \Fig{Fppm_trans}),
applies also to intermediate
scales. In \Fig{Fpspec_allo_Hyp3b} we show how the magnetic power spectrum
shows a bump at intermediate wavenumbers 
(i.e.\ the wide maximum at wavenumbers $k < 20 k_1$)
travelling to the left toward smaller wavenumbers.
In order to check whether or not the speed of the spectral bumps in the
different runs shown in \Figs{Fpspec_pm_kin_Hyp2b}{Fpspec_allo_Hyp3b} also decreases with
decreasing magnetic diffusivity we plot in \Fig{Fpkpeak_comp} the position
of the secondary bump in the power spectrum
as a function of time.  It is
evident that the slope, which signifies the speed of the bump, decreases
with decreasing magnetic diffusivity. A reasonable fit to this migration
is given by
\EQ
k_{\max}^{-1}=\alpha_{\rm trav}(t-t_{\rm sat}),
\EN
where the parameter $\alpha_{\rm trav}$ characterises the speed at which
this secondary bump travels. The values obtained from fits to various
runs are also listed in \Tab{Tsum}. Following arguments given by Pouquet
\ea (1976), $\alpha_{\rm trav}$ obtained in this way is actually a
measure of the $\alpha$ effect. The decrease of $\alpha_{\rm trav}$ with
decreasing magnetic diffusivity adds to the evidence that
the $\alpha$ effect is quenched in a $R_{\rm m}$ dependent fashion.

%\begin{figure}[t!]\centering\includegraphics[width=0.5\textwidth]{fig/pkpeak_comp.ps}\caption{

The $R_{\rm m}$ dependence of $\alpha_{\rm trav}$ and $S/(2M)$, both indicative
of inverse energy transfer, seems to be in conflict with the usual idea of
an inverse cascade where such transfers happen on a dynamical time scale
(e.g., Verma 2001). The only reasonable conclusion from this seems to be that
the (nonlocal!) inverse transfer discussed here is in fact distinct from
the usual inverse cascade. In other words, the present simulations do
not support the concept of an inverse cascade in the {\it usual} (local) sense.

\subsection{Magnetically driven turbulence}

\begin{figure}[t!]\centering\includegraphics[width=0.5\textwidth]{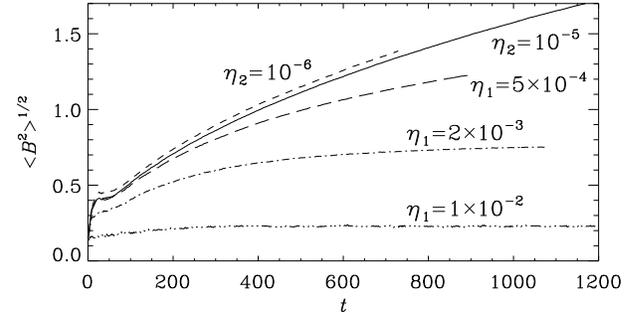}\caption{
Evolution of magnetic energy in runs with magnetic driving.
Forcing occurs in the induction equation around wavenumber $k_{\rm f}=5$.
Runs with hyperdiffusivity are denoted by $\eta_2$ and those with ordinary
diffusivity by $\eta_1$.
}\label{Fpn_bforc12}\end{figure}

In the cases considered so far, the turbulence is driven
by a forcing term in the momentum equation.
Another way to inject energy into the system is by adding an irregular
external electromotive force
{\boldmath${\cal E}$}$_{\rm ext}(\xx,t)$ to the system.
The induction equation (\ref{Induction-AAA}) is then replaced by
\EQ
  \frac{\partial\AAA}{\partial t}
  = \uu\times\BB + \eta\nabla^2\AAA
    + \mbox{\boldmath${\cal E}$}_{\rm ext}(\xx,t) .
\EN
This is in some ways reminiscent of turbulence driven by magnetic
instabilities (e.g.\, the magnetorotational instability, buoyancy or
kink instabilities). The direct injection of energy into the induction
equation was also considered by Pouquet \ea (1976), for example, who
assumed the forcing to be helical.
The results of such a run are shown
in \Fig{Fpn_bforc12}. Note that the magnetic energy evolves to larger
values as the magnetic diffusivity is decreased.
This is because the injection of magnetic energy is primarily balanced
by magnetic diffusion, which becomes weaker as the magnetic diffusivity
is decreased.
In the runs with second
order hyperdiffusivity (denoted by $\eta_2$) the magnetic energy evolves
to even larger values.
What is more important, however, is the time scale on which saturation
is achieved. This time scale is the resistive one, suggesting again that
the transport of magnetic energy across the spectrum from small to large
scales is a resistively limited process.

%\begin{figure}[t!]\centering\includegraphics[width=0.5\textwidth]{fig/pn_bforc12.ps}\caption{

\section{Evolution of the large scale magnetic helicity}
\label{Shelevol}

\subsection{Magnetic helicity flux from global mean-field models}
\label{mhelsun}

The only information available to date about the magnetic helicity of the
sun is from surface magnetic
fields, and these data are incomplete. Vector magnetograms of active regions
show negative (positive) current helicity on the northern (southern)
hemisphere (Seehafer 1990, Pevtsov, Canfield, \& Metcalf 1995, Bao \ea 1999, Pevtsov \&
Latushko 2000). From local measurements one can only obtain the current
helicity, so nothing can be concluded about magnetic helicity. Typically,
however, current and magnetic helicities agree in sign, but the current
helicity is stronger at small scales.

Berger \& Ruzmaikin (2000) have estimated the flux of
magnetic helicity from the solar surface. They discussed $\alpha$-effect
and differential rotation as the main agents facilitating the loss of
magnetic helicity.  Their results indicate that the flux of magnetic
helicity due to rotation and the actual magnetic field obtained from
solar magnetograms
is negative (positive) in the northern (southern) hemisphere.
Thus, the sign agrees with that of the current helicity obtained using
vector magnetograms. Finally, Chae (2000) estimated the magnetic
helicity flux based on counting the crossings of pairs of flux tubes.
Combined with the assumption that two nearly aligned flux tubes are
nearly parallel, rather than antiparallel, his results again
suggest that the magnetic helicity is negative (positive) in the
northern (southern) hemisphere.

What is curious here is that even though different scales are involved
(small and intermediate scales in active regions and coronal mass
ejections, and large scales in the differential rotation), there is
as yet no evidence that the helicity spectrum of the sun reverses sign at
some scale.

We now discuss the magnetic helicity properties of
mean-field $\alpha\Omega$ dynamos in a sphere.
There are different reasons why looking at mean-field dynamos can
be interesting in connection with helical turbulence. Firstly,
magnetic helicity involves one wavenumber factor less than the
magnetic energy and tends therefore to be dominated by the large
scale field. Magnetic helicity should therefore be describable
by mean-field theory. Secondly, mean-field dynamos easily allow
the spherical geometry to be taken into account. This is mainly
because such models can be only two-dimensional. The resulting field
geometry does in some ways resemble the solar field. We recall,
however, that at present there is no satisfactory model that
explains both the equatorward migration of magnetic activity as
well as the approximate antiphase between radial and azimuthal field
(${\overline B}_r{\overline B}_\phi<0$
during most of the cycle; see Stix 1976), and is still compatible
with the observed differential rotation ($\partial\Omega/\partial
r>0$ at low latitudes).
The sign of differential rotation does not however affect the sign of the
magnetic helicity. In fact, the velocity does never generate magnetic
helicity, it contributes only to the flux of magnetic helicity;
see \Eqs{dABdt}{helcons}.

\begin{table}[htb]\caption{
Properties of mean-field dynamos for different signs of $\alpha$
and radial shear, $\Omega'=\partial\Omega/\partial r$.
All of these models are problematic when applied to the sun.
 Case~(iii),
where $\alpha_0>0$ and $\Omega'<0$, satisfies two important
constraints (butterfly diagram with equatorward migration,
antiphase between radial and azimuthal field), but it is
incompatible with the helioseismology data which
indicate that at least at low latitudes $\Omega'>0$.
}\vspace{12pt}\centerline{\begin{tabular}{c||c|c}
             & $\Omega'<0$   & $\Omega'>0$  \\
\hline
$\alpha_0<0$ & Case (i)               & Case (ii)               \\
             & helioseismology not OK & {\bf helioseismology OK}\\
             & butterfly not OK       & {\bf butterfly OK}     \\
             & {\bf phase rel.\ OK}   & phase rel.\ not OK     \\
& $H_{\rm N}$ and $Q_{\rm N}$ {\bf negative}& $H_{\rm N}$ and $Q_{\rm N}$ {\bf negative}\\
\hline
\hline
$\alpha_0>0$ & Case (iii)             & Case (iv)               \\
             & helioseismology not OK & {\bf helioseismology OK}\\
             & {\bf butterfly OK}     & butterfly not OK        \\
             & {\bf phase rel.\ OK}   & phase rel.\ not OK      \\
& $H_{\rm N}$ and $Q_{\rm N}$ positive& $H_{\rm N}$ and $Q_{\rm N}$ positive\\
\label{Taosum}\end{tabular}}\end{table}

For the axisymmetric mean field, $\meanBB$, of an $\alpha\Omega$ dynamo
we calculate the gauge-invariant relative magnetic helicity of
Berger \& Field (1984) for the northern hemisphere (indicated by N),
\EQ
H_{\rm N}=\int_{\rm N}(\meanAA+\meanAA_{\rm P})
\cdot(\meanBB-\meanBB_{\rm P})\;\dd V,
\label{BF84integral}
\EN
where $\meanBB_{\rm P}=\nab\times\meanAA_{\rm P}$ is a potential reference
field inside the northern hemisphere, whose radial component agrees
with $\meanBB$ (see Bellan 1999 for a review). This implies that
${\overline A}_\phi={\overline A}_{{\rm P}\phi}$ on the boundary
of the northern hemisphere (both on the outer surface and the equator).

In spherical coordinates an axisymmetric magnetic field can be expressed
in the form $\meanBB={\overline B}_\phi\vec{\hat\phi}+\nab\times{\overline A}_\phi\vec{\hat\phi}$.
The potential reference field is purely poloidal. Therefore, the dot product
of the toroidal components in the integral \eq{BF84integral} is simply
\EQ
H_{\rm N}^{\rm(tor)}=\int_{\rm N}({\overline A}_\phi
+{\overline A}_{{\rm P}\phi}){\overline B}_\phi\;\dd V.
\EN
For the dot products of the poloidal components of the integral we express
the magnetic vector potential in the form
$\meanAA={\overline A}_\phi\vec{\hat\phi}+\nab\times{\overline C}_\phi\vec{\hat\phi}$,
where $\vec{\hat\phi}\cdot\nab\times\nab\times({\overline C}_\phi\vec{\hat\phi})={\overline B}_\phi$.
We can then write the integrand for the poloidal component as
\EQA
(\nab\times{\overline C}_\phi\vec{\hat\phi})\cdot\nab
\times({\overline A}_\phi-{\overline A}_{{\rm P}\phi})\vec{\hat\phi}
={\overline B}_\phi({\overline A}_\phi-{\overline A}_{{\rm P}\phi})
\nonumber\\
-\nab\cdot[\meanAA\times({\overline A}_\phi-{\overline A}_{{\rm P}\phi})\vec{\hat\phi}].
\ENA
(Note that ${\overline C}_{{\rm P}\phi}=0$, as the potential reference field
is purely poloidal.)
The integral over the divergence term vanishes because
${\overline A}_\phi={\overline A}_{{\rm P}\phi}$ on the boundary, so we have
\EQ
H_{\rm N}^{\rm(pol)}=\int_{\rm N}{\overline B}_\phi
({\overline A}_\phi-{\overline A}_{{\rm P}\phi})\;\dd V,
\EN
and hence
\EQ
H_{\rm N}=2\int_{\rm N}{\overline A}_\phi{\overline B}_\phi\;\dd V.
\label{BF84integral2}
\EN
The time evolution of $H_{\rm N}$ can be obtained from the
mean-field dynamo equation (e.g.\ Krause \& R\"adler 1980),
which gives
\EQ
\dd H_{\rm N}/\dd t=2S-Q_{\rm N},
\label{dHmdt_sph}
\EN
where $S=\int_{\rm N}\meanemf\cdot\meanBB\;\dd V$ was already
defined in \Eq{Sexpression},
$\meanemf=\alpha\meanBB-\eta_{\rm T}\mu_0\meanJJ$ is the e.m.f.\
of the mean field, and
$Q_{\rm N}$ is the integrated magnetic helicity flux out of the northern
hemisphere (i.e.\ through the outer surface and the equator), with
\EQ
Q_{\rm N}=-2\oint[(\meanemf+\meanUU\times\meanBB)
\times{\overline A}_\phi\vec{\hat\phi}]\cdot\dd\SSS,
\EN
where $\meanUU=r\sin\theta\,\Omega\vec{\hat\phi}$ is the mean
toroidal flow. It will be of interest to split $Q_{\rm N}$ into
the contributions from the outer surface integral, $Q_{\rm NS}$,
and the equatorial plane, $Q_{\rm Eq}$, so
$Q_{\rm N}=Q_{\rm NS}+Q_{\rm Eq}$.

In \Tab{Taosum} we summarise the key properties of typical $\alpha\Omega$
dynamo models for different signs of $\alpha$ and 
$\Omega'\equiv\partial\Omega/\partial r$ (cf.\ Stix 1976).
Here we have also indicated the sign of the magnetic helicity flux
that leaves from the northern hemisphere, $Q_{\rm N}$.
In all the models presented here the sign of $Q_{\rm N}$ agrees with
that of $H_{\rm N}$, which also agrees with the sign of $\alpha$.
This is because, averaged over one cycle,
$\bra{\dd H_{\rm N}/\dd t}_{\rm cyc}$ vanishes, so the sign of
$\bra{Q_{\rm N}}_{\rm cyc}$ must agree with that of
$\bra{S}_{\rm cyc}$ which,
in turn, is determined by the sign of $\alpha$; see \Eq{dHmdt_sph}.

What remains unclear, however, is whether the negative current helicity
observed in the northern hemisphere is indicative of small or large
scales, and hence whether the sign of $\alpha$ is positive (case iv)
or negative (case ii), respectively.
Only in case (iv) there would be the possibility of shedding magnetic
helicity of the opposite sign; see \Sec{Sremoval} below.
In case (iii), $H_{\rm N}$ is also positive,
but here $\Omega'<0$, which does not apply to the sun.
In case (ii), on the other hand, $H_{\rm N}$ is
negative, so shedding of negative magnetic helicity would act as a loss.

Our goal is to estimate the amount of magnetic helicity that is to be
expected for more realistic models of the solar dynamo. We also need to know which
fraction of the magnetic field takes part in the 11-year cycle.
In the following we drop the subscript N on $H$ and $M$.
For an oscillatory dynamo, all three
variables, $H$, $M$, and $Q_{\rm Joule}$ vary in an oscillatory
fashion with a cycle frequency $\omega$ of magnetic energy
(corresponding to 11 years for the sun -- not 22 years), so
we estimate
$|\dd H/\dd t|\la\omega|H|$ and $Q_{\rm Joule}\la\omega M$, which,
by arguments analogous to those used in \Sec{Subsec-cancellation},
leads to the inequality
\EQ
|H|/(2\mu_0 M)\leq \ell_{\rm skin},
\label{H-M-skin}
\EN
where $\ell_{\rm skin} = (2\eta/\omega)^{1/2}$ is the skin depth,
now associated with the oscillation frequency $\omega$.
Thus, the maximum magnetic helicity that
can be generated and dissipated during one cycle is characterised
by the length scale $|H|/(2\mu_0 M)$, which has to be less than
the skin depth $\ell_{\rm skin}$.

\begin{table*}[htb]\caption{
Results from a mean-field dynamo model in a spherical shell with
outer radius $R$ and inner radius $0.7R$, with either purely latitudinal shear
`(lat)' or purely radial shear `(rad)'.
The values of $\alpha$ given here (in units of $\eta_{\rm T}/R$) lead
to dynamos that are only slightly supercritical.
The cycle period, $T_{\rm cyc}$, is given in non-dimensional form
and would correspond to $22\yr$ for $\eta_{\rm T}=2.4\times10^{11}\cm^2/\s$
(for the models with latitudinal shear), or $3.7\times10^{11}\cm^2/\s$
(for the models with radial shear).
The range of ${\overline B}_{\rm pole}/{\overline B}_{\rm belt}$
that is consistent with observations, $(2\ldots5)\times10^{-4}$,
is highlighted in bold face.
Both $M_{\rm N}$ and $H_{\rm N}$ vary with the dynamo cycle period;
$H_{\rm N}^{\max}$ and $\bra{H_{\rm N}}$ refer to the maximum and time
averaged values of $H_{\rm N}$, respectively.
$\bra{Q_{\rm NS}}T_{\rm cyc}$ is the outwards magnetic helicity flux
through the northern surface, while $\bra{Q_{\rm Eq}}T_{\rm cyc}$ is the
magnetic helicity flux through the equator into the southern hemisphere.
}\vspace{12pt}\centerline{\begin{tabular}{ccccccccccc}
$C_\alpha$ & $C_\Omega$  &
$\displaystyle{{\overline B}_{\rm pole}\over{\overline B}_{\rm belt}}$ &
$\displaystyle{H_{\rm N}^{\max}\over2\mu_0M_{\rm N}R}$ &
$\displaystyle{\bra{H_{\rm N}}\over2\mu_0M_{\rm N}R}$ &
$\displaystyle{\bra{Q_{\rm NS}}T_{\rm cyc}\over{\overline B}_{\rm belt}^2R^4}$ &
$\displaystyle{\bra{Q_{\rm Eq}}T_{\rm cyc}\over{\overline B}_{\rm belt}^2R^4}$ &
$\displaystyle{T_{\rm cyc}\eta_{\rm T}\over R^2}$
\vspace{1mm}
\\
\hline
$2\times10^{-1}$ & $1\times10^5$ (lat) &$\bf1\times10^{-3}$&$1\times10^{-3}$ & $4\times10^{-4}$&$6\times10^{-4}$&$7\times10^{-6}$&0.030\\%ao2la-1
$6\times10^{-2}$ & $3\times10^5$ (lat) &$\bf3\times10^{-4}$&$3\times10^{-4}$ & $1\times10^{-4}$&$2\times10^{-4}$&$2\times10^{-6}$&0.030\\%ao2la-1.3
$2\times10^{-2}$ & $1\times10^6$ (lat) & $1\times10^{-4}$ & $1\times10^{-4}$ & $4\times10^{-4}$&$6\times10^{-5}$&$7\times10^{-7}$&0.030\\%ao2la-2
\hline
$2\times10^{-0}$ & $3\times10^3$ (rad) &$\bf8\times10^{-4}$&$5\times10^{-3}$ & $5\times10^{-3}$&$6\times10^{-4}$&$2\times10^{-3}$&0.047\\%ao2ra-0
$6\times10^{-0}$ & $1\times10^4$ (rad) &$\bf2\times10^{-4}$&$2\times10^{-3}$ & $1\times10^{-3}$&$3\times10^{-4}$&$5\times10^{-4}$&0.047\\%ao2ra-1
$2\times10^{-1}$ & $3\times10^4$ (rad) & $8\times10^{-5}$ & $5\times10^{-4}$ & $5\times10^{-4}$&$7\times10^{-5}$&$2\times10^{-4}$&0.047\\%ao2ra-2
\label{Ttimescale}\end{tabular}}\end{table*}

For $\eta$ we have to use the Spitzer resistivity which is
proportional to $T^{-3/2}$ ($T$ is temperature), so $\eta$ varies between
$10^4\cm^2/\s$ at the base of the convection zone to about $10^7\cm^2/\s$
near the surface layers and decreases again in the solar atmosphere.
Using $\omega=2\pi/(11\yr)=2\times10^{-8}\s^{-1}$
for the relevant frequency at which $H$ and $M$ vary we have
$\ell_{\rm skin}\approx10\km$ at the bottom of the convection zone
and $\ell_{\rm skin}\approx300\km$ at the top.
This needs to be compared with the value $|H|/(2\mu_0 M)$ obtained from
dynamo models.

\begin{figure}[t!]\centering\includegraphics[width=0.5\textwidth]{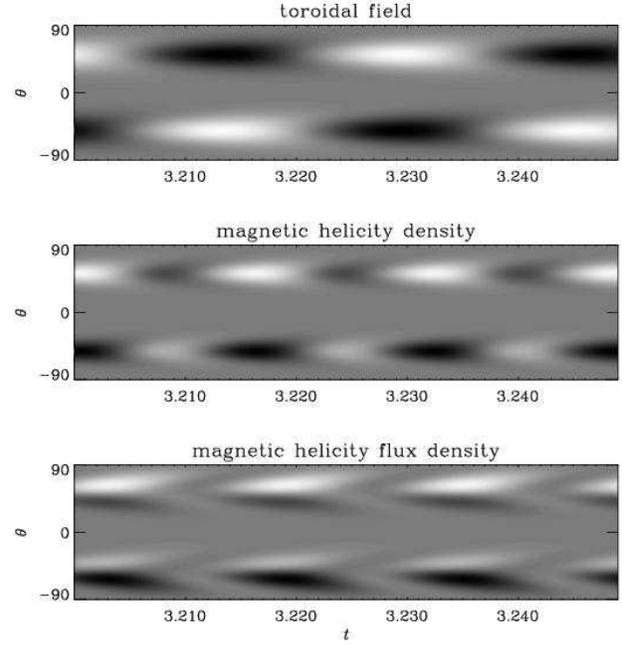}\caption{
Butterfly diagrams of the toroidal field from a mean field model,
compared with butterfly diagrams of the magnetic helicity density near
the surface and magnetic helicity flux due to differential rotation at
the surface. Note that the local magnetic helicity density
is most of the time positive (light grey) in the northern hemisphere and
negative (dark grey) in the southern hemisphere.
The magnetic helicity flux shows a similar variation, but there is a
certain degree of cancellation taking place at any moment in time.
$C_\alpha=0.175$, $C_\Omega=1\times10^5$ (lat).
}\label{Frbut_ao2}\end{figure}

In order to see how the condition (\ref{H-M-skin}) is met by
mean-field models, we present the results from a typical $\alpha\Omega$
model of the solar dynamo.
Although mean-field theory
has been around for several decades, the helicity aspect has only recently
attracted attention. In the proceedings of a recent meeting devoted specifically
to this topic (Brown, Canfield, \& Pevtsov 1999) helicity was discussed extensively
also in the context of mean-field theory. However, the possibility of
helicity reversals at some length scale was not addressed at the time.
We recall that in the Babcock--Leighton approach it is mainly the latitudinal
differential rotation that enters. We also note that, although the latitudinal
migration could be explained by radial differential rotation, meridional
circulation is in principle able to drive meridional migration even when
the sense of radial differential rotation would otherwise be wrong for driving
meridional migration (Durney 1995, Choudhuri, Sch\"ussler, \& Dikpati 1995,
K\"uker, R\"udiger, \& Schultz 2001).
We therefore proceed with the simplest possible model in spherical geometry.
An example of the resulting butterfly diagrams of toroidal field,
magnetic helicity and magnetic helicity flux is shown in \Fig{Frbut_ao2}.

%\begin{figure}[t!]\centering\includegraphics[width=0.5\textwidth]{fig/rbut_ao2.eps}\caption{

The main parameters that can be changed in the model are $\alpha_0$ (we assume that
$\alpha=\alpha_0\cos\theta$, where $\theta$ is colatitude), $\eta_{\rm T}$ (=const),
and the magnitude of the angular velocity, $\Omega_0$, in the expression
\EQ
\Omega=\Omega_0\left[1-0.038{P_3^1(\cos\theta)\over\sin\theta}
-0.004{P_5^1(\cos\theta)\over\sin\theta}\right],
\label{latshear}
\EN
which is a fit that resembles closely the surface angular velocity of the sun
(R\"udiger 1989). What matters for the axisymmetric dynamo is only
the pole--equator difference, $\Delta\Omega$, which is about a third
of $\Omega_0$.  Our model is made dimensionless by putting $R=\eta_{\rm
T}=1$.  The governing non-dimensional parameters are $C_\alpha=\alpha_0
R/\eta_{\rm T}$ and $C_\Omega=\Delta\Omega R^2/\eta_{\rm T}$, where
$\Delta\Omega=\Omega_{\max}-\Omega_{\min}$. This leads
to the observables ${\overline B}_{\rm pole}$, by which we mean the radial field at
the pole, which is about 2 gauss for the sun, and ${\overline B}_{\rm belt}$,
by which we mean
the azimuthal field in the main sunspot belts.
The solar value of ${\overline B}_{\rm belt}$
is estimated to be around $4\ldots8\kG$, based on equipartition arguments and on
the total emergent flux
during one cycle.\footnote{For the sun the total emergent flux per cycle
is about $10^{24}\Mx$. Assuming that the toroidal field is distributed over a
meridional surface of $\sim(20-50)\times500\Mm^2\approx(1-3)\times10^{20}\cm^2$,
this amounts to an estimated toroidal field of $3-10\kG$.}
Thus, we expect ${\overline B}_{\rm pole}/{\overline B}_{\rm belt}$ to be in
the range $(2\ldots5)\times10^{-4}$.

From the model results shown in \Tab{Ttimescale} we see that,
once ${\overline B}_{\rm pole}/{\overline B}_{\rm belt}$
is in the range consistent with observations,
(highlighted in bold face), $\bra{H_{\rm N}}/(2\mu_0M_{\rm N}R)$
is around $(1\ldots3)\times10^{-4}$ for models with latitudinal shear
using \Eq{latshear}.
Given that $R=700\Mm$ this means that
$H_{\rm N}/(2\mu_0M_{\rm N})\approx70\dots200\km$, which would be compatible
with $\ell_{\rm skin}$ only near the top of the convection zone.

In units of ${\overline B}_{\rm belt}^2R^4$, the magnetic helicity flux
through the northern hemisphere is about $(2\ldots5)\times10^{-4}$.
Using typical parameters for the sun, ${\overline B}_{\rm belt}=4\kG$ and
$R=700\Mm$, we have ${\overline B}_{\rm belt}^2R^4=4\times10^{50}\Mx^2$.
Thus, our models predict magnetic helicity fluxes of about
$(1\ldots2)\times10^{47}\Mx^2$ for a full 22 year magnetic cycle,
or half of that for the 11 year activity cycle.
This is comparable to the values reported by Berger \& Ruzmaikin (2000),
but less that those given by DeVore (2000).
Significant magnetic helicity fluxes through the equator are
only found if angular velocity contours cross the equatorial plane,
i.e.\ in the models with radial differential rotation.

If the observed negative magnetic helicity flux on the northern hemisphere is
associated with the large scale field, it would suggest that $\alpha_0<0$.
This corresponds to case (ii), where magnetic helicity flux would contribute
as an additional loss of large scale magnetic energy. On the other hand,
if the observed negative magnetic helicity flux is associated with the small
scale field, it could remove excess magnetic helicity of the opposite
sign, which may act as a source of large scale magnetic energy;
see \Sec{Sremoval}. This
corresponds to case (iv) where $\alpha_0>0$.

\subsection{Balance equations}

The models with imposed positive helical forcing at small length scale
show clearly that there is a production of equal amounts of positive
and negative magnetic helicity at small and large scales, respectively
(or vice versa
if the helicity of the forcing is negative). As time goes on, more and more
positive and negative magnetic helicity builds up on small and large scales.
Of course,
magnetic helicity always implies magnetic energy as well (this is related to the
realisability condition; see \Sec{Srealisability} and Moffatt 1978), and
eventually the value of the small scale magnetic energy will have reached
the value of the kinetic energy at that scale, so the dynamo
reaches saturation. This happens first at small scales, and if that was
where the dynamo stops, the resulting field would be governed by
small scales (Kulsrud \& Anderson 1992).
However, this is not what happens in the majority of
the numerical experiments, and probably also not in real astrophysical
bodies like the sun.

Before we go on, we need to address the question of opposite helicities
in the two hemispheres of astrophysical dynamos.
For reasons of symmetry, the total magnetic
helicity is always close to zero. It is then convenient
to think of the two hemispheres as two systems coupled with
each other through an open boundary (the equatorial plane).
The equation of magnetic helicity conservation
has an additional flux term, $Q$, on the right hand side, so it reads
[cf.\ \Eq{dABdt}]
\EQ
{\dd H\over\dd t}=-2\eta C-Q,
\EN
where $H$ is the magnetic helicity, $C$ the current helicity, $\eta$
the microscopic magnetic diffusivity, and $Q$ the magnetic helicity flux
integrated 
over the surface of the given subvolume (i.e.\ a hemisphere in this
case).\footnote{$Q$ is not to be confused with the rate of Joule
heating, $Q_{\rm Joule}$.}
We are interested in the evolution of the large scale field
$\meanBB$, and we refer to the magnetic helicity of this large scale or
mean field as $H_{\rm m}$.

In the absence of boundaries we have
$H_{\rm m}=\int\meanAA\cdot\meanBB\,\dd V$,
where $\meanBB=\nab\times\meanAA$
and overbars denote a suitable average that commutes with the
curl operator and satisfies the Reynolds rules (in particular, the average
of fluctuations vanishes and the average of an average gives the same;
see Krause \& R\"adler 1980). With open boundaries, however,
$\int\meanAA\cdot\meanBB\,\dd V$ is no longer gauge invariant. For the
present discussion we only need the fact that it is possible to define
gauge-invariant versions of magnetic helicity; see Berger \& Field (1984)
for the three-dimensional case and BD01 (or \Sec{Sopenbc}) for the
one-dimensional case that is relevant to horizontally averaged fields.

Splitting the field into contributions from mean and fluctuating
components we can write
\EQ
{\dd H_{\rm m}\over\dd t}+{\dd H_{\rm f}\over\dd t}
=-2\eta C_{\rm m}-2\eta C_{\rm f}-Q_{\rm m}-Q_{\rm f}.
\label{dHtotdt}
\EN
This equation only becomes useful if we can make appropriate simplifications.
BD01 considered the case where the small scale field
has already reached saturation and the large scale magnetic energy continues
to build up further until it too reaches saturation. 

If the large scale field is
nearly fully helical we have $M_{\rm m}\approx\half k_{\rm m}|H_{\rm m}|$,
where $k_{\rm m}$ is the typical wavenumber of the mean field.
Furthermore, if the small scale field is nearly fully helical, too, we have
$M_{\rm f}=\half k_{\rm f}|H_{\rm f}|$, where $k_{\rm f}$ is the typical
wavenumber of the fluctuating field, or the forcing wavenumber, as in
B01.
Analogous relations link $C$ to $H$.
 The relative ordering of the ratios of large and
small scale magnetic energies relative to magnetic and current helicities
is given by
\EQ
{|H_{\rm m}|\over|H_{\rm f}|}\approx
{k_{\rm f}\over k_{\rm m}}{M_{\rm m}\over M_{\rm f}}\approx
\left({k_{\rm f}\over k_{\rm m}}\right)^2{|C_{\rm m}|\over|C_{\rm f}|}.
\EN
Thus, if $k_{\rm m}\ll k_{\rm f}$, as is usually
the case when there is some degree of scale separation,
and  $M_{\rm m} \approx M_{\rm f}$, we have
$|H_{\rm m}|\gg |H_{\rm f}|$ and
$|C_{\rm m}|\ll|C_{\rm f}|$, so we are left with
\EQ
{\dd H_{\rm m}\over\dd t}\approx
-2\eta C_{\rm f}-2Q_{\rm m}-2Q_{\rm f},
\label{dHmdt}
\EN
or, using the above estimates between $M$, $H$ and $C$, and assuming,
for definiteness, that $H_{\rm m},C_{\rm m}>0$ and $H_{\rm f},C_{\rm f}<0$
(as expected in the northern hemisphere of the sun), we have
\EQ
{\dd M_{\rm m}\over\dd t}\approx+2\eta k_{\rm m}k_{\rm f}
M_{\rm f}-2k_{\rm m}Q_{\rm m}-2k_{\rm m}Q_{\rm f}.
\label{dMmdt}
\EN
(The factor 2 in front of the $Q$s comes from the 1/2 in the
definition of $M$.)
We can apply this equation to the problem with open boundaries.

However before doing this, we recall that in the absence of any 
surface fluxes of helicity ($Q=0$) the assumption $M_{\rm m}=M_{\rm f}$ 
does not apply near saturation. Instead, steadiness
of $H$ immediately requires $C=C_{\rm m}+C_{\rm f}=0$, so
$|C_{\rm m}|=|C_{\rm f}|$, and hence
$M_{\rm m}=(k_{\rm f}/k_{\rm m})\,M_{\rm f}\gg M_{\rm f}$,
and $|H_{\rm m}| \gg |H_{\rm f}|$ for the final state. Before
this final state is reached we have from \Eq{dHtotdt}
\EQ
{\dd M_{\rm m}\over\dd t}\approx
+2\eta k_{\rm m}k_{\rm f}M_{\rm f}
-2\eta k_{\rm m}^2M_{\rm m},
\label{dMmdt2}
\EN
which has the solution (B01)
\EQ
M_{\rm m}\approx{k_{\rm f}\over k_{\rm m}}M_{\rm f}\,
\left[1-e^{-2\eta k_{\rm m}^2(t-t_{\rm s})}\right].
\label{Mm1}
\EN
This equation is referred to as the magnetic helicity constraint, which
is probably obeyed by all helicity-driven large scale dynamos.
Note, however, that no mean-field theory has been invoked here.
\EEq{Mm1} implies that full saturation is reached not earlier than after
a resistive time scale, $(\eta k_{\rm m}^2)^{-1}$.
We emphasise that this is the resistive time scale connected to the
{\it large} length scale $2\pi/k_{\rm m}$.
We also emphasise that this result is unaffected by turbulent or
other (e.g.\ ambipolar; see BS00) magnetic diffusion.

Coming back to the problem with open boundaries we can carry out a similar
analysis by relating $Q_{\rm m}$ to the mean field. Based on dimensional
arguments we may assume
\EQ
|Q_{\rm m}|=M_{\rm m}/(k_{\rm m}\tau_{\rm m}),
\EN
where $\tau_{\rm m}$ is some damping time for the large scale field. We
can express this in terms of an effective (turbulent-like) magnetic diffusion
coefficient, $\eta_{\rm m}$, for the large scale field, e.g.\ in the form
\EQ
2\eta_{\rm m}=(\tau_{\rm m} k_{\rm m}^2)^{-1}
\quad\mbox{(definition of $\eta_{\rm m}$)},
\EN
so \Eq{dMmdt} becomes
\EQ
{\dd M_{\rm m}\over\dd t}\approx
+2\eta k_{\rm m}k_{\rm f} M_{\rm f}
-2\eta_{\rm m} k_{\rm m}^2 M_{\rm m}
-2k_{\rm m}Q_{\rm f},
\EN
which has the solution
\EQ
M_{\rm m}\approx{2\eta k_{\rm f}M_{\rm f}-2Q_{\rm f}\over
2\eta_{\rm m}k_{\rm m}}\,
\left[1-e^{-2\eta_{\rm m} k_{\rm m}^2(t-t_{\rm s})}\right].
\label{Mm2}
\EN
We expect $\eta_{\rm m}$ to scale with the turbulent magnetic
diffusivity, $\eta_{\rm t}$, which describes the spreading of
the mean magnetic field.
This equation yields the time scale $(\eta_{\rm m}k_{\rm m}^2)^{-1}$, which is
indeed shorter than the resistive time scale, $(\eta k_{\rm m}^2)^{-1}$.
However, if $Q_{\rm f}=0$ in \Eq{Mm2}, the energy of the final field relative to
the small scale field is decreased by the factor
$(\eta k_{\rm f})/(\eta_{\rm m}k_{\rm m})\ll1$. This was the basic
result of BD01 where $|Q_{\rm f}|$ was found to be much smaller than
$|Q_{\rm m}|$. In other words, all the magnetic large scale helicity
loss occurred at large scales, which is of course the scale where we
wanted to generate the large scale field.

The above results have shown that magnetic helicity losses due to
large scale magnetic fields are not really a satisfactory solution to the problem
of generating strong large scale magnetic fields on fast enough time
scales. Rather, one requires that there should be a loss of 
magnetic helicity due to small scale magnetic fields, i.e.\
a non zero $Q_{\rm f}$. We may
assume that $Q_{\rm f}$ and $Q_{\rm m}$ have opposite sign, because both
current and magnetic helicities also have different signs for the contributions
from small and large scales. Looking back at \Eq{Mm2} is it clear that
the $Q_{\rm f}$ term in the numerator can in principle enhance and
even completely supersede the resistively limited `driver',
$2\eta k_{\rm f}M_{\rm f}$, provided $Q_{\rm f}<0$ on the northern
hemisphere.

Physically speaking, the role of small scale magnetic helicity flux is
to let the large scale field build up independently of any constraints
related to the small scales. To prevent then excessive build-up of
magnetic energy at small (and intermediate) scales, which diffusion
was able to do (albeit slowly), we need to have another more direct
mechanism that is independent of microscopic magnetic diffusion. In
order to investigate the problem of excess magnetic helicity at small
scales we have performed a series of numerical experiments by
Fourier filtering the field and removing in regular time intervals
the magnetic energy at the forcing and smaller scales.

\subsection{Removing excess magnetic helicity}
\label{Sremoval}

According to our working hypothesis, the reason
why the large scale field can only develop slowly is that the
build-up of the large scale field with one sign of magnetic
helicity requires the simultaneous build-up (and eventual
removal) of small scale field with opposite magnetic helicity.
Instead of waiting for ohmic field destruction one might expect
that also the explicit removal of magnetic energy at small scales would
be able to accomplish this goal.

\begin{figure}[t!]\centering\includegraphics[width=0.5\textwidth]{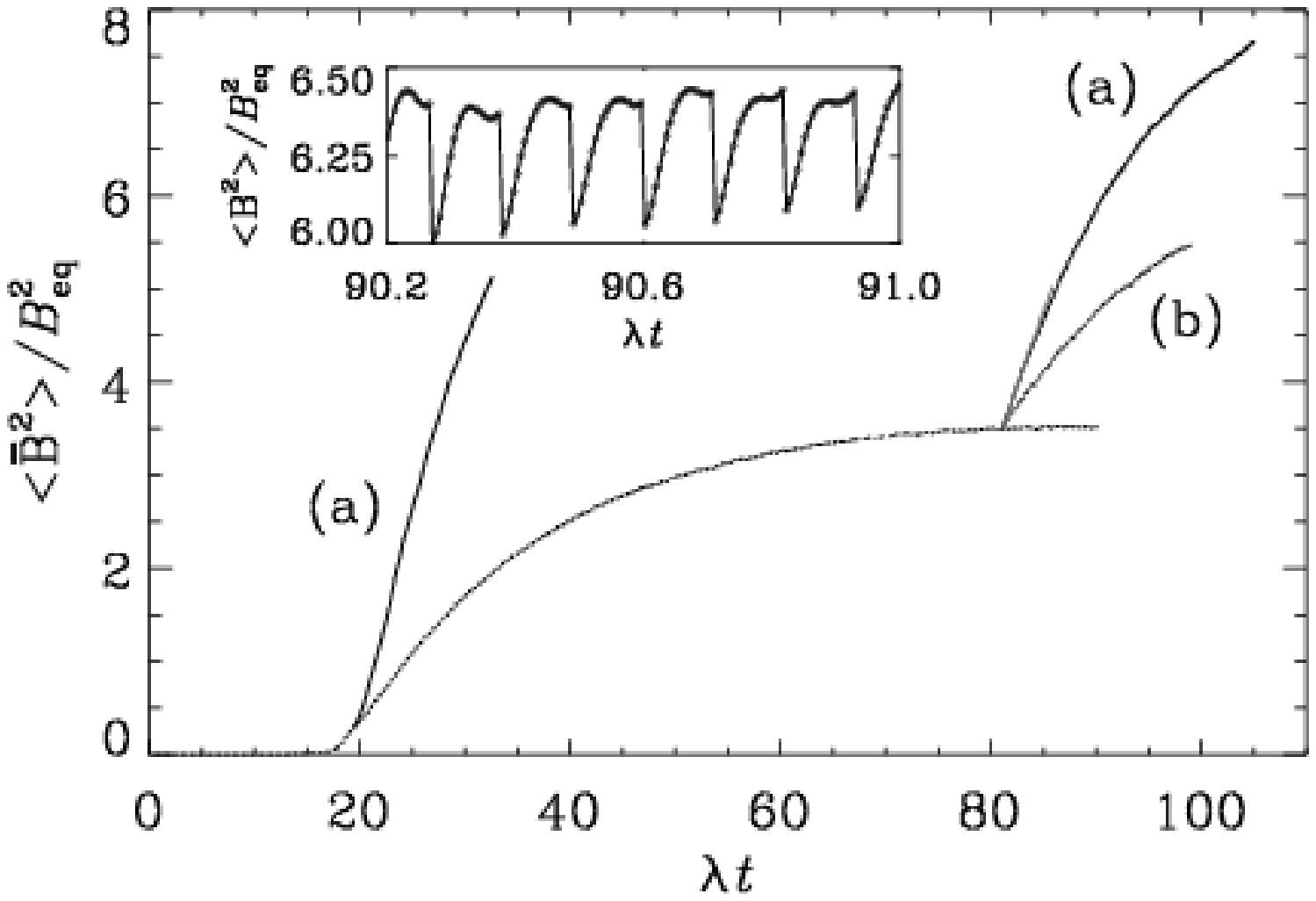}\caption{
The effect of removing small scale magnetic energy in regular time intervals
$\Delta t$ on the evolution of the large scale field (solid lines).
The dashed line gives the evolution of $\bra{\meanBB^2}$ for Run~3 of B01
(where no such energy removal was included) in units of
$B_{\rm eq}^2=\mu_0\rho_0\bra{\uu^2}$. The two solid lines show the evolution of
$\bra{\meanBB^2}$ after restarting the simulation from Run~3 of B01
at $\lambda t=20$ and $\lambda t=80$. Time is scaled with the kinematic
growth rate $\lambda$. The curves labelled (a) give the result for
$\Delta t=0.12\lambda^{-1}$ and those labelled (b) for $\Delta t=0.4\lambda^{-1}$.
The inset shows, for a short time interval, the
sudden drop and subsequent recovery of the total (small and large scale)
magnetic energy in regular time intervals.
}\label{Fpbmean}\end{figure}

A more formal argument is this. We use again \Eq{dHtotdt}, but without
ignoring the $\dd H_{\rm f}/\dd t$ term, and make the assumption that
large and small scale fields are fully helical, and ignore surface
terms, because boundary conditions are assumed periodic. Thus,
we have an additional $\dd M_{\rm f}/\dd t$ term in \Eq{dMmdt2},
which then reads
\EQ
{\dd M_{\rm m}\over\dd t}\approx
{k_{\rm m}\over k_{\rm f}}{\dd M_{\rm f}\over\dd t}
+2\eta k_{\rm m}k_{\rm f}M_{\rm f}
-2\eta k_{\rm m}^2M_{\rm m}.
\label{dMmdt2_ddt}
\EN
After small scale magnetic energy is removed, the right hand side
of \Eq{dMmdt2_ddt} is out of balance. If we ignored the
$\dd M_{\rm f}/\dd t$ term, the large scale magnetic energy would
decay proportional to $-2\eta k_{\rm m}^2M_{\rm m}$, because the
$2\eta k_{\rm m}k_{\rm f}M_{\rm f}$ term is still negligible after having
set $M_{\rm f}=0$. However, the $M_{\rm f}$ term begins to recover
immediately which gives rise to a significant contribution from the
$\dd M_{\rm f}/\dd t$ term. Both terms enter with the same sign.
If the magnetic Reynolds number is large,
i.e.\ if the recovery rate of $M_{\rm f}$ is faster than than the
microscopic decay time, we have a net gain. This is indeed borne
out by the simulations, where magnetic energy is removed above
the wavenumber $k=4$ every
$\Delta t=0.12\lambda^{-1}$ time units, labelled (a) in \Fig{Fpbmean},
and $\Delta t=0.4\lambda^{-1}$, labelled (b).
Here, $\lambda=0.07u_{\rm rms}k_{\rm f}$ is the kinematic
growth rate of the rms field strength. Shorter time intervals do not
give any further enhancement; the resulting curve for
$\Delta t=0.04\lambda^{-1}$ lies almost on top of the curve (a)
that starts at $\lambda t=80$.

%\begin{figure}[t!]\centering\includegraphics[width=0.5\textwidth]{fig/pbmean.ps}\caption{

We may thus conclude that the preferential removal of small scale
magnetic fields does indeed enhance large scale dynamo action. Earlier
experiments, which are not reported here, where $k_{\rm f}$ was larger,
and therefore the magnetic Reynolds number $u_{\rm rms}/\eta k_{\rm f}$
smaller, did not show such an enhancement. Note however that we still
do not really know whether the preferential loss of such small scale field
does operate in real astrophysical bodies. Simulations have so far
been unsuccessful in demonstrating this. Those presented in BD01 show
that most of the magnetic helicity is lost through the large scale field.
A possible reason may be that the large scale field geometry was not
optimal and that in other settings large scale losses are reduced.

The intermediate scales are probably those where most of the magnetic
helicity loss from the sun occurs. This argument is based in particular
on the sign of the flux of magnetic helicity on the solar surface: it is
found to be negative (positive) on the northern (southern) hemisphere
%(Seehafer 1990, Pevtsov \ea 1995, Bao \ea 1999, Pevtsov \& Latushko 2000, Chae 2000).
(see \S\ref{mhelsun} above).

According to conventional mean-field theory the sign of the $\alpha$-effect
should be positive in the northern hemisphere, so the sign of large scale
magnetic helicity and the flux thereof should be positive. A reversal of
the sign of magnetic helicity can then be expected on some scale smaller
than the
large scale of the dynamo wave ($100\mbox{--}300\Mm$). Thus, the helicity flux
occurs on the scale of active and bipolar regions as well as the scale of
coronal mass ejections, which is about $20\mbox{--}50\Mm$.
These scales may therefore well correspond to the intermediate scales
referred to above.
On the other hand, there is some uncertainty
as to whether the dynamo alpha in the northern hemisphere of the sun might
in fact be negative.
If the dynamo $\alpha$ is negative in the northern hemisphere,
the magnetic helicity loss on the sun would be of the same sign as that of
the large scale dynamo-generated field, so its scale would still belong to
the large scales. However, the solution to the helicity
problem would then probably not be related to magnetic helicity flux, because
then $Q_{\rm f}$ and $Q_{\rm m}$, as well as $H_{\rm m}$, would have the same sign,
so $Q_{\rm f}$ would constitute an additional loss term, which does hence
not accelerate the dynamo (BD01); see also
\Sec{Shelevol}.

%\begin{figure*}[t!]\centering\includegraphics[width=0.95\textwidth]{fig/curl_15b2.eps}\caption{
\begin{figure*}[t!]\centering\includegraphics[width=0.95\textwidth]{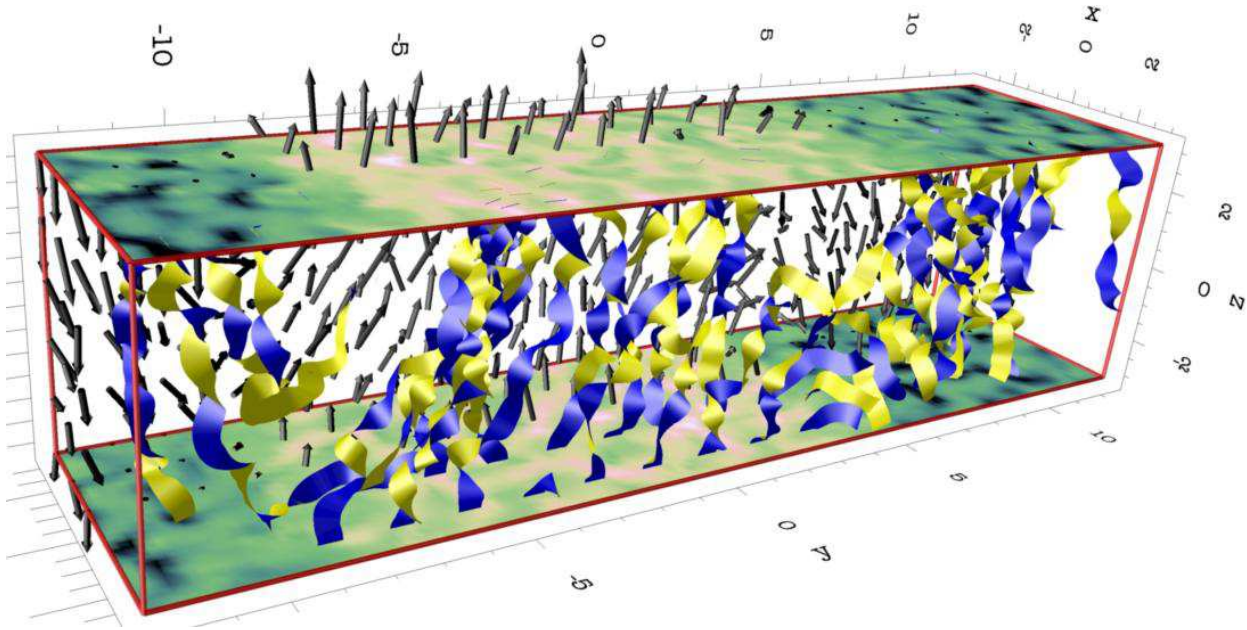}\caption{
Three-dimensional visualisation of the magnetic field for a simulation
where the box is four times longer in the $y$-direction.
Vectors in the back half of the box show the orientation of the locally
averaged field.
Ribbons in the front half indicate magnetic field lines; their twist
reflects the local torsality (or helicity) of the magnetic field
and has been enhanced artificially by factor of 15.
The vertical magnetic field component is indicated by bright (positive) or
dark (negative) colours in the top and bottom face of the box.
}\label{Fdxfigure}\end{figure*}

\subsection{Large aspect ratio simulations}

In order to assess the role of intermediate scales,
we now extend the simulations of BD01
with open boundaries to domains that are larger in one horizontal
coordinate direction compared to before.
We have already mentioned in
the introduction that the simulations of BD01
produced most of the magnetic helicity at large scale, i.e.\ the same
scale as that of the large scale field that we want to build up fast. A potential
problem with these simulations could be a lack of sufficient scale
separation. In particular, it is conceivable that intermediate scales
are not sufficiently well represented in these simulations.
The intermediate length scales
can be expected to be where most of the magnetic helicity of the
opposite sign resides (`opposite' is here meant to be
relative to the large scale field).

\begin{figure}[t!]\centering\includegraphics[width=0.5\textwidth]{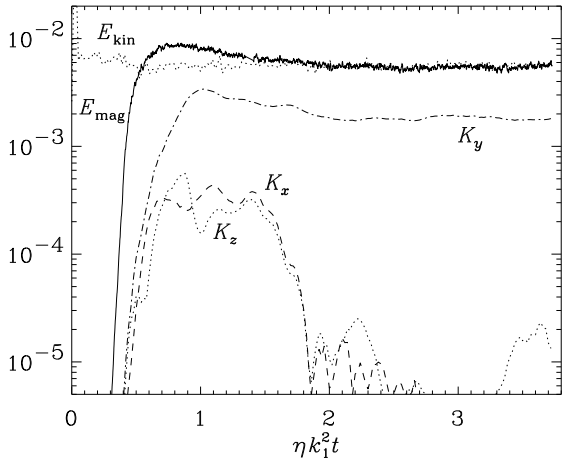}\caption{
Evolution of kinetic and magnetic energies. The magnetic energy
contained in the large scale field (averaged in the $x$ and $z$
directions, and denoted by $K_y$) is shown as a dash-dotted line.
The dotted and dashed curves denote the $z$- and $x$-dependent components,
respectively.
}\label{Fasp3t}\end{figure}

In order to study the possible role of the allowance of intermediate
length scales in the simulations we extend the simulations of BD01
to the case of larger aspect ratio. The hope is that the large scale
field will now develop in the coordinate direction that is longest,
but that most of the magnetic helicity flux occurs still at a similar
scale as before and that the sign of magnetic helicity and its flux
are now reversed relative
to the helicity at the scale of the mean field. In the simulations of
BD01 a box with unit aspect ratio was considered, and one of the
coordinate directions (the $z$-direction) was non-periodic and magnetic
helicity flux through these boundaries occurred. We have now extended this
study to the case where the box is four times larger in
one of the other directions (the $y$-direction).
A three-dimensional visualisation of
the resulting field is shown in \Fig{Fdxfigure}.
The $y$ component of the magnetic field on the $z$-boundaries shows a
clear large scale pattern with a length scale larger than the extent
of the domain in the $x$ direction.
The evolution of the magnetic and kinetic
energies of one such run is shown in \Fig{Fasp3t}.
We see that kinetic and magnetic energies are approximately in equipartition.
The large scale magnetic energy shows, as expected, variation in the
$y$ direction, i.e.\ the wave vector of the large scale field points
in the long direction of the box.

%\begin{figure}[t!]\centering\includegraphics[width=0.5\textwidth]{fig/asp3t.ps}\caption{

%\begin{figure}[t!]\centering\includegraphics[width=0.5\textwidth]{fig/pebm_comp.ps}\caption{
\begin{figure}[t!]\centering\includegraphics[width=0.5\textwidth]{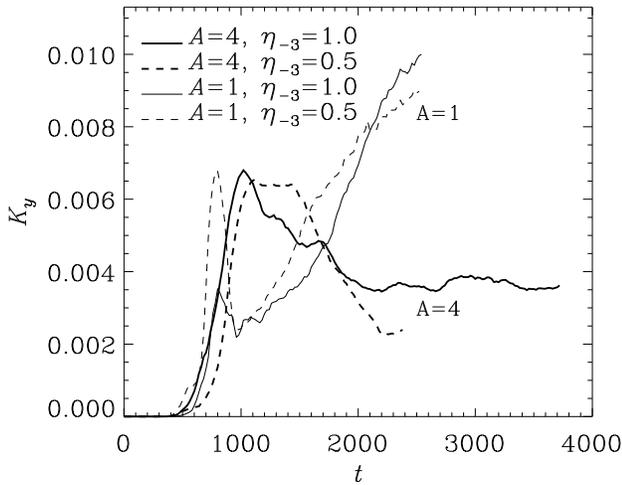}\caption{
Comparison of the evolution of the large scale magnetic field component
that varies in the $y$-direction (the direction in which the box is
longest) for runs with different aspect ratios, $A$, and different
values of $\eta_{-3}=\eta/10^{-3}$.
Note the development of a stronger large scale field for $A=1$ than for $A=4$.
}\label{Fpebm_comp}\end{figure}

\begin{table}[t!]\caption{
Summary of the main properties of the runs with vertical field boundary
condition. The parameter $q_j$ gives the fractional magnetic energy in
the mean field relative to the total magnetic field.
}\vspace{12pt}\centerline{\begin{tabular}{llllccccc}
 & $\!\!\!A\!\!\!$ & $\!\!\nu\!\!$ & $\!\!\eta\!\!$
&$\!\!u_{\rm rms}\!\!$&$\!\!b_{\rm rms}\!\!$
& $q_z\!\!$ & $\!\!q_x\!\!$ & $\!\!q_y\!\!$ \\
\hline
Vert~1 &$\!$ 1 $\!$&$\!$ 0.01  & 0.01   & 0.10 & 0.09 & 0.76 & 0.48 & 0.48 \\
Vert~2 &$\!$ 1 $\!$&$\!$ 0.005 & 0.005  & 0.16 & 0.14 & 0.59 & 0.45 & 0.45 \\
Vert~3 &$\!$ 1 $\!$&$\!$ 0.002 & 0.002  & 0.20 & 0.18 & 0.39 & 0.30 & 0.30 \\
Vert~4 &$\!$ 1 $\!$&$\!$ 0.002 & 0.001  & 0.19 & 0.17 & 0.24 & 0.20 & 0.20 \\
A1/10  &$\!$ 1 $\!$&$\!$ 0.02  & 0.001  & 0.11 & 0.12 & 0.61 & 0.47 & 0.47 \\
A1/05  &$\!$ 1 $\!$&$\!$ 0.02  &$0.0005\!\!$& 0.11 & 0.13 & 0.56 & 0.43 & 0.43 \\
A4/10  &$\!$ 4 $\!$&$\!$ 0.02  & 0.001  & 0.11 & 0.11 & 0.00 & 0.00 & 0.33 \\
A4/05  &$\!$ 4 $\!$&$\!$ 0.02  &$0.0005\!\!$& 0.11 & 0.12 & 0.02 & 0.02 & 0.19 \\
\label{T1}\end{tabular}}\end{table}

Different runs are compared in \Fig{Fpebm_comp}.
All the models with open boundaries have the property of developing
rapidly a mean field that varies in the $y$-direction, which is the
long direction. When the aspect ratio $A$ is small, however, the {\it final}
field is not that which varies in the $y$-direction, but instead in the
$z$-direction. This is seen from \Tab{T1} where we compare the
relative strengths of the mean field,
\EQ
q_j=K_j/M\equiv\bra{\meanBB}_j^2/\bra{\BB^2},\quad j=x,y,z
\EN
where $\bra{\meanBB}_j$ is the mean field averaged in the two directions
perpendicular to $j$ (cf.\ B01).
In all the cases with $A=1$, $q_z$ is larger than $q_x$ and $q_y$.
This is not the case when $A=4$: here $q_z$ and $q_x$ are very small.
Nevertheless, as the magnetic Reynolds number is increased, $q_y$
decreases (from 0.33 to 0.19 when $\eta$ is lowered from $10^{-3}$ to
$5\times10^{-4}$). Also, comparing $q_y$ in the $A=4$ case with
$q_z$ is the $A=1$ case, the mean field is clearly less strong.

Thus, we may conclude that larger aspect ratios favour large scale
magnetic field configurations that vary in the horizontal ($y$)
direction instead of the vertical ($z$) direction. However, at late
times the saturation field strength is still decreasing with decreasing
resistivity. It seems therefore worthwhile pursuing this work by studying
still larger aspect ratios and yet different field geometries.

\subsection{Fractional small scale magnetic helicity}
\label{Sfractional}

In practice the magnetic helicity will never be near 100\%; typical
values are around 3-5\% for convective and accretion disc turbulence
(Brandenburg \ea 1995, 1996). This has prompted Maron \& Blackman (2002)
to consider the effects of a forcing that has only fractional helicity
(i.e.\ the realisability condition for the velocity is not saturated).
They found that
there is a threshold in the degree of helicity above which (large-scale)
dynamo action is possible. However, these results were obtained at a
resolution of only $64^3$ meshpoints, and it is therefore possible,
that this threshold effect is really the result of an
increase of the critical magnetic Reynolds number above which
the large scale $\alpha$-type dynamo operates, and that the model
with that resolution has dropped below this critical value
for large scale dynamo action. Below we will estimate whether the dynamo
number was in fact large enough for this to happen.

The effect of fractional helicity of the
forcing on the evolution of the large scale magnetic
energy can be assessed in terms of the magnetic helicity constraint,
using a generalised form of \Eq{dMmdt2}. In the case of fractional
magnetic helicity the small and large scale magnetic helicities will
be less than expected based on the magnetic helicity constraint, so
\EQ
|C_{\rm f}|/k_{\rm f}\approx\epsilon_{\rm f}M_{\rm f}
\EN
and
\EQ
k_{\rm m}|H_{\rm m}|=|C_{\rm m}|/k_{\rm m}
\approx\epsilon_{\rm m}M_{\rm m},
\EN
where $\epsilon_{\rm f}\leq1$ and $\epsilon_{\rm m}\leq1$ denote the
degree of the resulting helicity on small and large scales, respectively.
Then, instead of \Eq{Mm1}, we have
\EQ
M_{\rm m}\approx{\epsilon_{\rm f}k_{\rm f}\over\epsilon_{\rm m}k_{\rm m}}M_{\rm f}\,
\left[1-e^{-2\eta k_{\rm m}^2(t-t_{\rm s})}\right],
\label{Mmfrac}
\EN
which is just $\epsilon_{\rm f}/\epsilon_{\rm m}$ times the expression on
the right hand side of \Eq{Mm1}.
Two things can happen: if the small scale field is only weakly helical,
i.e.\ $\epsilon_{\rm f}\ll1$, but still $\epsilon_{\rm m}\approx1$,
then the large scale magnetic energy
will be decreased. On the other hand, if the large scale magnetic field
is only weakly helical, i.e.\ $\epsilon_{\rm m}\ll1$,
but $\epsilon_{\rm f}\approx1$, then the large
scale magnetic energy will actually {\it increase} relative to
the prediction of \Eq{Mm1}. In practice, both
effects could happen at the same time, in which case the change of
the large scale magnetic energy will be less strong.

%\begin{figure}[t!]\centering\includegraphics[width=0.5\textwidth]{fig/pn_Partial1+8.eps}\caption{
\begin{figure}[t!]\centering\includegraphics[width=0.5\textwidth]{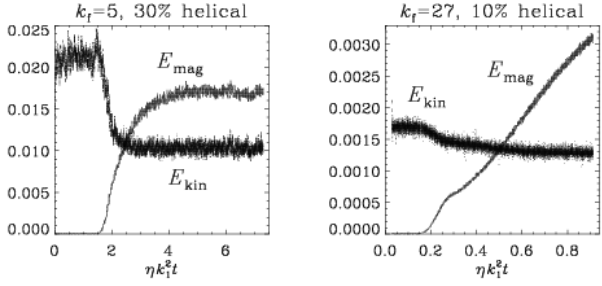}\caption{
Evolution of kinetic and magnetic energies in two runs with fractional
helicity. On the left hand panel, $k_{\rm f}=5$ and $\eta=2\times10^{-3}$,
whilst on the right hand panel, $k_{\rm f}=27$ and $\eta=2\times10^{-4}$.
}\label{Fpn_Partial1+8}\end{figure}

In \Fig{Fpn_Partial1+8} we show the results of two runs with fractional
helicity forcing. In both cases the magnetic energy exceeds the kinetic
energy by $k_{\rm f}/k_{\rm m}$ multiplied by an `efficiency factor',
$\epsilon_{\rm f}/\epsilon_{\rm m}$. In the first case we have
$k_{\rm f}/k_{\rm m}=5$ and $\epsilon_{\rm f}/\epsilon_{\rm m}=0.3$,
giving super-equipartition by about 1.5. This is roughly in agreement with
\Fig{Fpn_Partial1+8}. In the second case we have $k_{\rm f}/k_{\rm m}=27$
and $\epsilon_{\rm f}/\epsilon_{\rm m}=0.1$, giving super-equipartition by
a factor of
about 2.7. Although the run in the second panel of \Fig{Fpn_Partial1+8}
has not been run for a full magnetic diffusion time (which is here ten
times longer than in the first case), it seems clear that strong
super-equipartition is still possible.

In the framework of $\alpha^2$ dynamo theory (Moffatt 1978,
Krause \& R\"adler 1980), the large scale field is described
by \Eq{dmeanAAdt}. In a periodic domain with minimum wavenumber $k_1$
the condition for large scale dynamo action is that
the dynamo number, $C_\alpha=\alpha/(\eta_{\rm T}k_1)$, exceeds unity.
Standard estimates are $\alpha=-\onethird\tau\bra{\oo\cdot\uu}$
and $\eta_{\rm t}=\onethird\tau\bra{\uu^2}$ (e.g., Moffatt 1978),
where $\tau$ is the correlation time.
For strongly helical turbulence with forcing at a particular wavenumber
$k_{\rm f}$, we have
$|\bra{\oo\cdot\uu}|\approx \epsilon_{\rm f}k_{\rm f}\bra{\uu^2}$, so
\EQ
C_\alpha
={|\alpha|/k_1\over\eta+\eta_{\rm t}}
={|\alpha|/(\eta_{\rm t}k_1)\over1+R_{\rm m}^{-1}}
=\epsilon_{\rm f}k_{\rm f}/\iota k_1,
\EN
where $\iota=1+R_{\rm m}^{-1}$ is a correction factor involving the
magnetic Reynolds number, defined here as $R_{\rm m}=\eta_{\rm t}/\eta$.
For the Runs shown in \Fig{Fpn_Partial1+8}, and with the numbers given
in the previous paragraph, the values of $C_\alpha$
are 1.4 and 2.7, respectively. Thus, the dynamo number is supercritical.
This is consistent with the clear growth of the large scale field
seen in the simulations (\Fig{Fpn_Partial1+8}).
Maron \& Blackman (2002) found that for $k_{\rm f}/k_1=5$ the critical
value of $\epsilon_{\rm f}$ is around 0.5. This corresponds to
$C_\alpha=2.5/\iota\approx2$, which should still be supercritical.
However, given that only modest resolution was used, the estimate
for $\alpha$ may have been too optimistic.

%\begin{figure}[t!]\centering\includegraphics[width=0.5\textwidth]{fig/pn_perf.eps}\caption{
\begin{figure}[t!]\centering\includegraphics[width=0.5\textwidth]{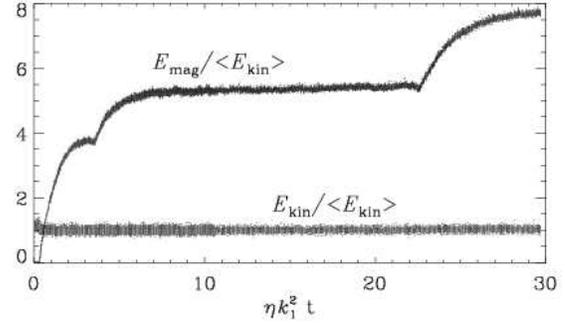}\caption{
Evolution of kinetic and magnetic energies in a run with perfectly
conducting boundaries in the $z$-direction.
Note that the magnetic energy evolves in stages, similar to \Fig{Fpmm}.
}\label{Fpn_perf}\end{figure}

\subsection{Fractional large scale magnetic helicity}

We have seen that the large scale field strength will decrease if the
small scale field is not fully helical. However, according to \Eq{Mmfrac},
the large scale magnetic field strength should increase if the large scale
field were less helical. In order to demonstrate that this is true,
we show in \Fig{Fpn_perf} the results of a run where the forcing
is still 100\% helical, but the large scale field is not fully helical. The latter has
been achieved by adopting perfectly conducting boundary conditions,
in which case it is no longer possible to have fully helical Beltrami waves.
These calculations are otherwise similar to those presented in BD01,
where the effects of nonperiodic boundary conditions was considered.

In \Fig{Fpn_perf} we see another remarkable feature that was already seen
in \Fig{Fpmm}: the large scale magnetic energy grows in stages. Indeed,
looking at the actual magnetic field pattern one sees that the number of
nodes in plots of the mean field decreases with time, just like in \Fig{Fpmm}.

\subsection{Position of the secondary peak}

\begin{figure}[t!]\centering\includegraphics[width=0.5\textwidth]{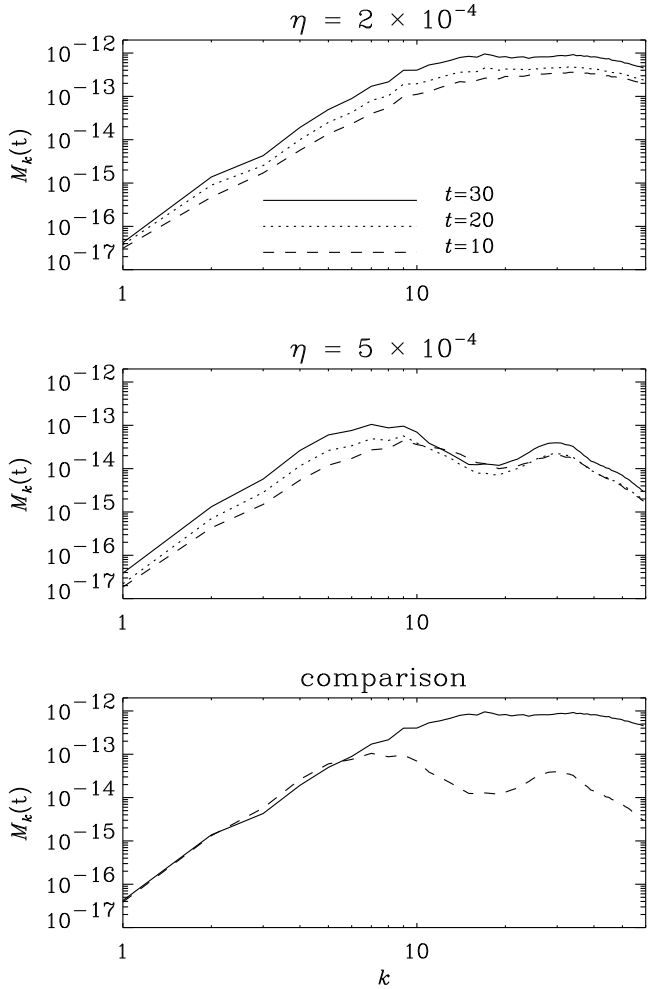}\caption{
Scale separation in runs with different resistivity (first and second
panel). Higher resistivity results in larger scale separation (third panel).
}\label{Fmkt_comp}\end{figure}

In the original Run~6 of B01 there
was a field of intermediate scale that
grew fastest at $k=7$. This wavenumber was found
to be in agreement with $k_{\max}=|\alpha|/(2\eta_{\rm T})$, which is what
one expects for an $\alpha^2$ dynamo. Here
$\eta_{\rm T}=\eta+\eta_{\rm t}$ is the sum of microscopic and turbulent
magnetic diffusivity and $\alpha$ represents the $\alpha$ effect that is
caused by the helical small scale velocity. In the simulation, the only
parameter that can be changed to make $|\alpha|/(2\eta_{\rm T})$ vary is the
microscopic magnetic Reynolds number. In order to check our prediction regarding $k_{\max}$,
we reduce the value of $\eta$
from $\eta=10^{-3}$ (as in B01) to $\eta=2\times10^{-4}$. We find that now
the fastest growing mode is at $k_{\max}\approx15$, which is, on a logarithmic scale, almost
indistinguishable
from the forcing wavenumber, $k_{\rm f}=30$; see \Fig{Fmkt_comp}.
At larger scales, $k<7$, the exponential growth is equally
fast in all modes, independent of the (microscopic) value of $\eta$.

%\begin{figure}[t!]\centering\includegraphics[width=0.5\textwidth]{fig/mkt_comp.ps}\caption{

The reason for decreasing scale separation with increasing value of
$R_{\rm m}$ is relatively easy to see. According to the dispersion relation
for the $\alpha^2$ dynamo in an infinite domain (e.g., Moffatt 1978)
the wavenumber of the fastest growing mode is
\EQ
k_{\max}=|\alpha|/2\eta_{\rm T}
=|\alpha|/2\iota\eta_{\rm t}
=\epsilon_{\rm f}k_{\rm f}/2\iota.
\EN
So, unless $R_{\rm m}$ is small, in which case $\iota=1+R_{\rm m}^{-1}$
can be large, $k_{\max}$ will be around $k_{\rm f}/2$ giving a scale
separation of 1:2. However, for small values of $\epsilon_{\rm f}$,
the scale separation can be much larger.

In the high $R_{\rm m}$ limit, a scale separation of just a factor
of 2 is hardly visible during
the kinematic evolution of the field at that scale. Once the dynamo is
in the nonlinear regime, i.e.\ once the small scale field has saturated,
$\alpha$ will be quenched and $k_{\max}$ must decrease further.

\begin{figure}[t!]\centering\includegraphics[width=0.5\textwidth]{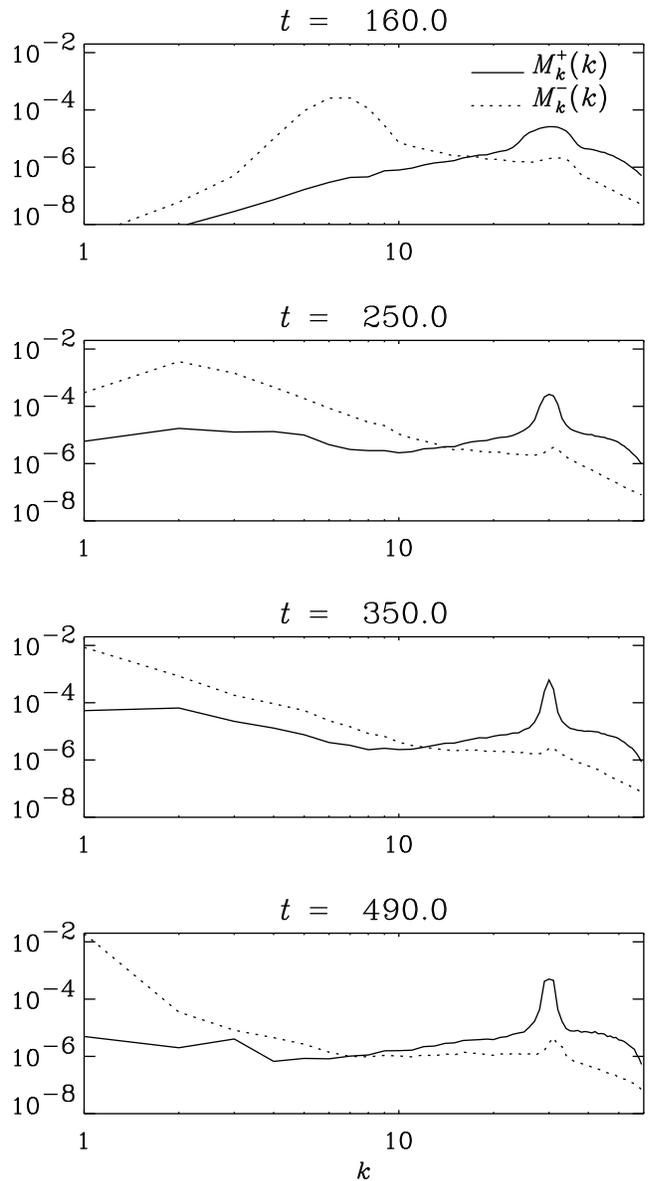}\caption{
Decomposition of magnetic field into positive and negative helical parts.
Note that the magnetic energy at the forcing scale has increased
between $t=250$ and $t=350$. The magnetic energy at large scales
is dominated by helicity with negative sign (dashed curves).
}\label{Fphel_dec_spec_Run6}\end{figure}

In order to see that the sign of the magnetic helicity is indeed
different at small and large scales we split the field into right
and left handed parts, $\BB=\BB^++\BB^-$
(see \Sec{Srealisability} and Christensson, Hindmarsh, \& Brandenburg
2001), and show in \Fig{Fphel_dec_spec_Run6} the magnetic energy spectra
of $\BB^+$ and $\BB^-$, $M_k^+$ and $M_k^-$, respectively.
Inverse transfer is only seen in the $M_k^-$ spectrum; the $M_k^+$
spectrum is dominated by the forcing scale.

%\begin{figure}[t!]\centering\includegraphics[width=0.5\textwidth]{fig/phel_dec_spec_Run6.ps}\caption{

\section{Dynamos with shear}
\subsection{Evidence for magnetic Reynolds number dependence}

Finally, we turn attention to the case of dynamos with shear.
The presence of shear allows for an additional induction effect that is
responsible for producing strong toroidal field from a poloidal (cross-stream)
field component. According to a result from mean-field theory such dynamos
can possess oscillatory solutions (e.g.\ Parker 1979).
In a recent study by Brandenburg, Bigazzi, \& Subramanian (2001, hereafter
BBS01), where a velocity shear of the form $u_y(x)\propto\sin x$ was
applied to helically driven turbulence in a periodic box, oscillatory
solutions were confirmed also for non-mean field dynamos.
The outstanding question, however,
is whether the resistive time scale, which is now known to affect
the saturation
phase of the dynamo (B01, BBS01), also affects the cycle period. This question
could not be answered conclusively in BBS01, because only one run
with one value of the magnetic Reynolds number was considered.

In the meantime we have accumulated data from another run that had twice
the value of the magnetic Reynolds number. The flow is driven by a forcing,
which consist of two components: a term varying sinusoidally in
the $x$-direction with wavenumber $k=1$ driving the shear and a term
consisting of random Beltrami waves with wavenumber $k_{\rm f}=5$. As
in BBS01, the ratio of the magnitudes of the two components of the
forcing function is about 1:100, resulting in a shear velocity that is
in the final state about 50 times larger than the poloidal rms velocity
of the turbulence. This is large enough so that we can expect
to be in the so-called `$\alpha\Omega$ regime'
where oscillatory solutions are the rule.

\begin{figure}[t!]\centering\includegraphics[width=0.5\textwidth]{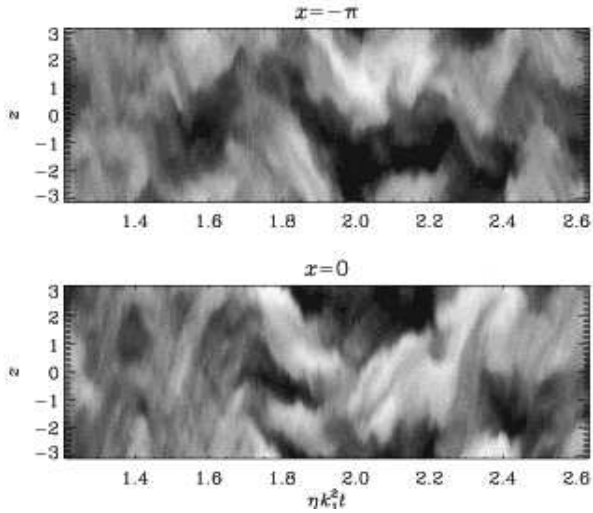}\caption{
Butterfly diagram of ${\overline B}_y$, continuation from the
run of BBS01, but with $\eta=2\times10^{-4}$.
Run~(iii) of \Tab{Tao2}.
}\label{Fpbutter6b}\end{figure}

Corresponding butterfly diagrams of the toroidally averaged toroidal
field components are shown in \Fig{Fpbutter6b}. The two diagrams are
for two different $x$-positions,
$x=\mp\pi$ and at $ x=0$,
where the shear $S=\dd u_y^{(0)}/\dd x$ takes on its negative and positive
extremum, respectively.
The simulation was restarted from
the run of BBS01 at the time $t=6\,000$ and run until $t=13\,000$ with
$\eta$ reduced by a factor of $2.5$ with respect to BBS01. It
turns out that the resulting magnetic field suffers a major disturbance
after the magnetic diffusivity is reduced by just a factor of 2. More
importantly, a clear pattern of a migratory dynamo wave is now almost absent.
Instead, the toroidal flux pattern appears to just wobble up and down in
the $z$ direction.

%\begin{figure}[t!]\centering\includegraphics[width=0.5\textwidth]{fig/pbutter6b.eps}\caption{

One possible explanation for this result is that the dynamo wave was
just very close to the threshold between oscillatory and nonoscillatory
behaviour. This is not very likely however: the estimates of BBS01
indicated that the dynamo numbers based on shear,
$C_\Omega=S/(\eta_{\rm T}k_1^2)$, is between 40 and
80, whilst the total dynamo number (${\cal D}=C_\alpha C_\Omega$) is between
10 and 20 (see BBS01), and hence $C_\alpha=\alpha/(\eta_{\rm T}k_1)\approx0.25$.
Thus, shear dominates strongly
over the $\alpha$-effect ($C_\Omega/C_\alpha$ is between 150 and 300), which
is typical for $\alpha\Omega$-type behaviour (i.e.~oscillations) rather
than $\alpha^2$-type
behaviour which would start when $C_\Omega/C_\alpha$ is below about 10
(e.g.\ Roberts \& Stix 1972).

%\begin{figure}[t!]\centering\includegraphics[width=0.5\textwidth]{fig/pbutter7b.eps}\caption{
\begin{figure}[t!]\centering\includegraphics[width=0.5\textwidth]{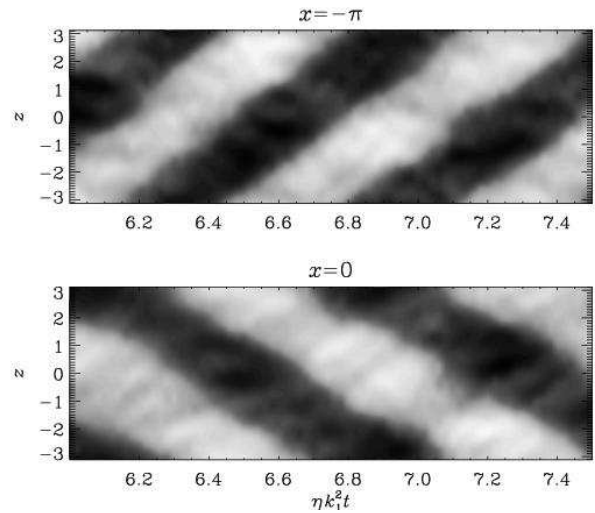}\caption{
Butterfly diagram of ${\overline B}_y$, continuation from the
run of BBS01, but with $\eta=10^{-3}$ (five times larger than in \Fig{Fpbutter6b}).
Run~(i) of \Tab{Tao2}.
}\label{Fpbutter7b}\end{figure}

Another possibility is that an oscillatory and migratory dynamo
is still possible, but it takes longer for the system to settle onto
this solution. Unfortunately the present simulations have become
prohibitively expensive in terms of computer time that this run cannot
be prolonged by much further at this time.
Instead, we have calculated a model with larger magnetic diffusivity
and found cycle periods that are now somewhat shorter than before
($700...800$ compared to $1000...2000$ in BBS02), but in diffusive
units they are similar ($0.7...0.8$ compared to $0.5...1.0$
in BBS02); see \Fig{Fpbutter7b}.
This suggests that the period in this oscillatory dynamo
is still controlled by the microscopic magnetic diffusivity.
A more detailed comparison of various parameters with the other
run presented above and with that of BBS01 is given in \Tab{Tao2}.

\begin{table}[t!]\caption{
Summary of the main properties of the three-dimensional
simulations with shear. Here, $\eta/(c_{\rm s}k_1)$ is
the magnetic diffusivity in units of the sounds speed and
the wavenumber of the domain,
and $\omega_{\rm cyc}=2\pi/T_{\rm cyc}$ is the
cycle frequency. In Run~(iii) there is no clear cycle visible.
}\vspace{12pt}\centerline{\begin{tabular}{lccc}
Run &  (i) & (ii) & (iii) \\
\hline
$\eta/(c_{\rm s}k_1)$ & $10^{-3}$ & $5\times10^{-4}$ & $2\times10^{-4}$ \\
$\nu/\eta$ & 5 & 10 & 25 \\
$\bra{\uu^2}^{1/2}/(\eta k_1)$ & 30 & 80 & 200 \\
$\bra{\meanUU^2}^{1/2}/(\eta k_1)$ & 600 & 1200 & 2500 \\
$\bra{\bb^2}/B_{\rm eq}^2$ &  4 & 6 & 20 \\
$\bra{\meanBB^2}/B_{\rm eq}^2$ & 20 & 30 & 60 \\
$(\bra{{\overline B}_x^2}/\bra{{\overline B}_y^2})^{1/2}$ & 0.018 & 0.014 & 0.008 \\
$\mu_0\bra{\meanJJ\cdot\meanBB}/\bra{\meanBB^2}$ & 0.11 & 0.06 & 0.014 \\
$\omega_{\rm cyc}/(\eta k_1^2)$ &  8\ldots9 & 6\ldots12 & $\ge10?$ \\
\label{Tao2}\end{tabular}}\end{table}

\subsection{Can magnetic helicity ride with the dynamo wave?}

\begin{figure}[t!]\centering\includegraphics[width=0.5\textwidth]{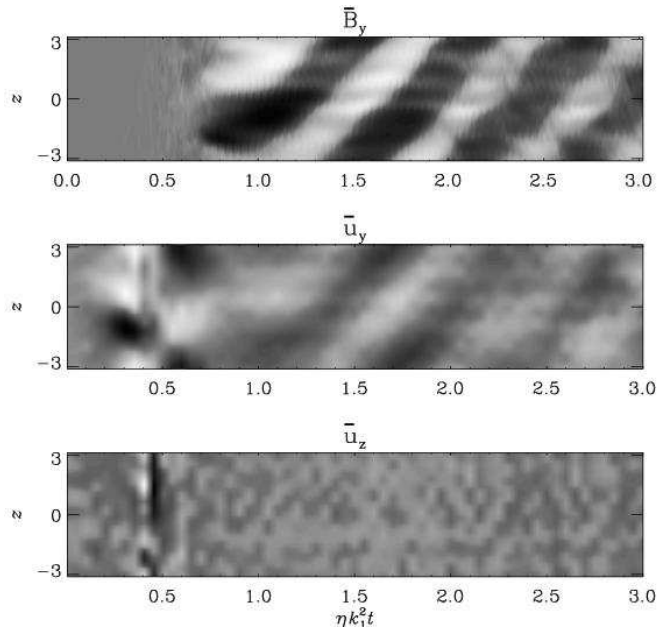}\caption{
Butterfly diagram of ${\overline B}_y$ for the run of BBS01
with $\eta=5\times10^{-4}$, compared with the corresponding diagrams
of ${\overline u}_y$ and ${\overline u}_z$ for the same time interval.
Note the presence of a `torsional' oscillation in ${\overline u}_y$,
i.e.\ a superposition of the driven shear flow (second panel),
but the absence of any similar pattern in ${\overline u}_z$
(last panel).
Run~(ii) of \Tab{Tao2}.
}\label{Fpbutter_uu+bb}\end{figure}

We now want to address the question whether magnetic helicity can
vary in space and time such that it is actually unchanged in the frame
of the dynamo wave as it propagates. The background of these suggestions
is as follows. Although the sign of the magnetic helicity in each hemisphere
remains the same (and is opposite in the other hemisphere), its value still
changes somewhat as the cycle proceeds. It is conceivable that even a small
fractional variation may prove incompatible with the Parker-type
migratory dynamo operating on a dynamical time scale.
One possibility worth checking is that the magnetic helicity pattern
could propagate together with the dynamo wave.
In that case the magnetic helicity would be conserved in a Lagrangian
frame moving with the wave, but not in an Eulerian frame.
If this transport is due to advection,
we would then expect systematic
sign changes of $u_z$ in a space-time diagram, which does not seem
to be the case however; see \Fig{Fpbutter_uu+bb}.

%\begin{figure}[t!]\centering\includegraphics[width=0.5\textwidth]{fig/pbutter_uu+bb.eps}\caption{

What does seem interesting, however, is the presence of what is usually
(in the context of solar physics) referred to as a `torsional' oscillation
(Howard \& LaBonte 1980). The toroidal flow pattern clearly traces the
toroidal magnetic field pattern. More recently Howe \ea (2000) found another shorter
period of 1.5 years which has been seen in helioseismological data.
If this can indeed also be interpreted as the result of a similar magnetic
field pattern (which is as yet undetected) then this would suggest the
presence of multiple periods in the solar dynamo wave
(cf.\ Covas, Tavakol, \& Moss 2001).

In the discussion above we had mainly bulk motions in mind that would
result in a transport of magnetic helicity. This does not need to be the
case because magnetic helicity can also be transported along twisting
magnetic field lines. This was used in
the approach by Vishniac \& Cho (2001), who proposed a dynamo effect based
on a non-vanishing divergence of the magnetic helicity flux.
If magnetic helicity were to ride with the dynamo wave, then this
could perhaps correspond to the anticipated magnetic helicity flux.
Unfortunately, in the parameter
regime currently accessible to simulations the effect proposed by Vishniac \& Cho
has not yet been confirmed (Arlt \& Brandenburg 2001). We may therefore conclude
that transport or advection of magnetic helicity by meridional flows remains
a possibility, but has not as yet been verified numerically.

\subsection{Helicity loss in the presence of surface shear}
\label{Snear_surface_shear}

The calculations of BD01 had the shortcoming that large scale shear was
absent. Of course, shear does not produce net magnetic helicity, but it
can lead to a spatial separation which could be particularly useful if
there are surfaces. Large scale shear in the equatorial plane would wind up a
poloidal field and hence would lead to magnetic helicity of opposite sign
in the northern and southern hemispheres. In helically forced turbulence
simulations, the various effects have only been studied in isolation:
shear and helicity in BBS01, surface shear but no helicity in Arlt \&
Brandenburg (2001), and different helicities in the northern and southern
hemispheres in Brandenburg (2001b).

Surface shear may allow for the direct production and transport of magnetic helicity
of different signs in the two hemispheres on a dynamical time scale.
Consider a differentially rotating star permeated by an initially uniform
magnetic field parallel to the rotation axis.
The differential rotation will wind up the magnetic field
in different directions in the two hemispheres, where the field lines
describe oppositely oriented screws. Such additional production and transport
of magnetic helicity on a dynamical time scale may be important for the magnetic
helicity problem. On the other hand, for this idea to be relevant we have to
have a pre-existing large scale poloidal magnetic field, so this issue does
not seem to address the problem of resistively limited field saturation.

In order to find out whether large scale surface shear is important
we have considered a model similar
to Run~C2 of Arlt \& Brandenburg (2001), i.e.\ with finite shear in a
layer of thickness $2\pi$ and a quiescent (non-turbulent) halo with no
shear outside.
As before, the shear is given by $u_y(x)$, while the boundary to the halo
is at $z = \pm\pi$.
 However, in contrast to the nonhelical calculations of
Arlt \& Brandenburg, we have now included a helical forcing that is negative
(positive) above (below) the midplane.
The results are in many respects similar to the
runs of BD01 with a halo: around the time of saturation there is a strong
negative burst of small scale magnetic helicity together
with positive small scale current helicity. The
resulting magnetic helicity fluxes both out of the turbulent domain
as well as upwards on the two boundaries are fluctuating about zero.
Thus, the addition of shear to the halo runs of BD01 seems
to have only little effect on the evolution of helicity fluxes.

\begin{figure}[t!]\centering\includegraphics[width=0.5\textwidth]{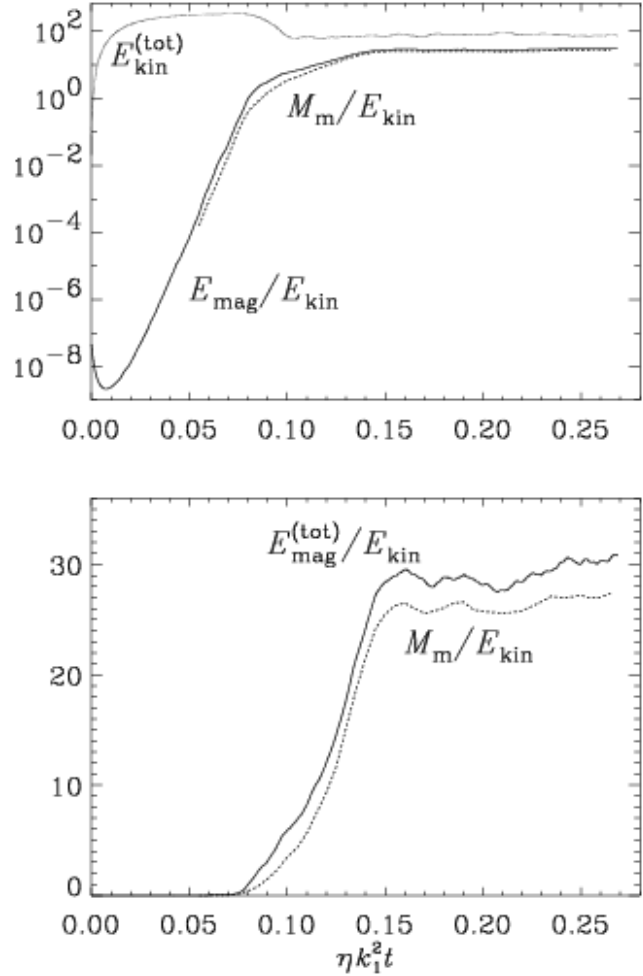}\caption{
Semi-logarithmic and double-linear plots showing the
evolution of kinetic and magnetic energies (dotted and solid lines),
compared with the magnetic energy of the azimuthally averaged mean field.
All quantities are normalised by the kinetic energy of the cross-stream
motions. $E_{\rm kin}^{\rm(tot)}$ refers to the total kinetic energy including
the shear. The ratio of the kinematic growth rate to the resistive
time at large scales is here $\lambda/\eta k_1^2=160$.
}\label{Fpbym_list}\end{figure}

\begin{figure*}[t!]\centering\includegraphics[width=0.95\textwidth]{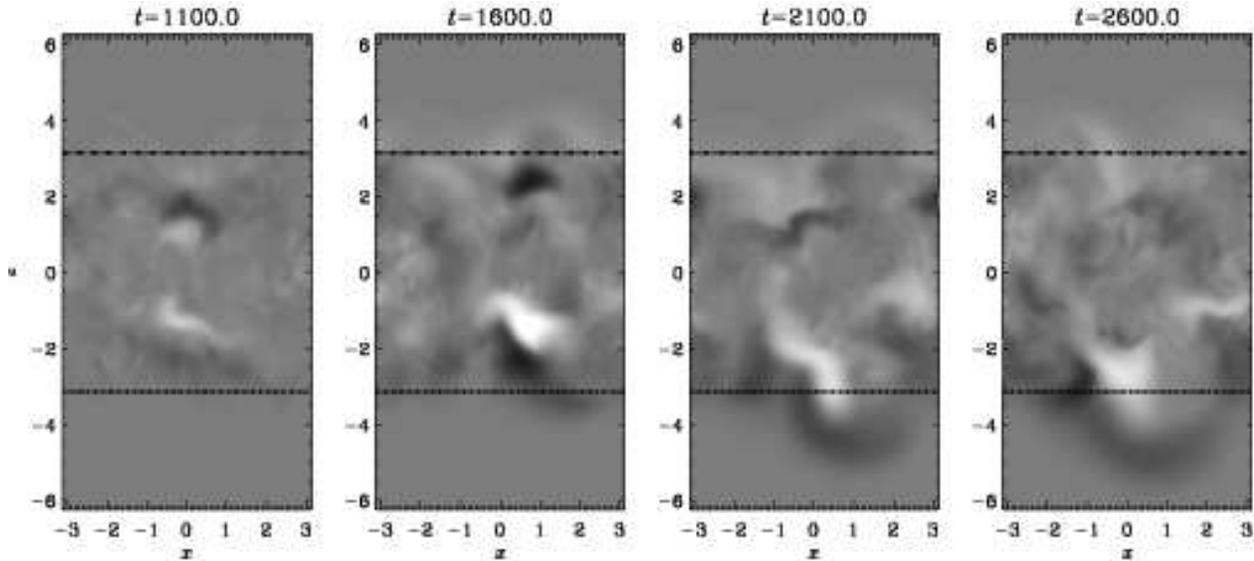}\caption{
Loss of magnetic field through the upper and lower boundaries.
Images of the mean toroidal field are shown at different times.
Shear is positive (negative) at $x=0$ ($x=\pm\pi$).
$\eta=10^{-4}$ and $\nu=5\times10^{-3}$.
The lines at $z=\pm\pi$ mark the boundaries of the turbulent slab.
}\label{Fppbym_aft}\end{figure*}

The evolution of the magnetic and kinetic energies is shown in \Fig{Fpbym_list}.
After about 0.08 resistive times the dynamo has entered the saturation
phase which is then completed after 0.15 resistive times altogether.
This is now much faster than in the case of periodic boundaries.
\Fig{Fppbym_aft} shows images of the mean toroidal magnetic field.
Especially near the lower boundary of the disc surface (dash-dotted line)
one sees the ejection of magnetic structures.

%\begin{figure}[t!]\centering\includegraphics[width=0.5\textwidth]{fig/pbym_list.ps}\caption{

%\begin{figure*}[t!]\centering\includegraphics[width=0.95\textwidth]{fig/ppbym_aft.eps}\caption{

%\begin{figure}[t!]\centering\includegraphics[width=0.5\textwidth]{fig/pbutter_ggravs5.eps}\caption{
\begin{figure}[t!]\centering\includegraphics[width=0.5\textwidth]{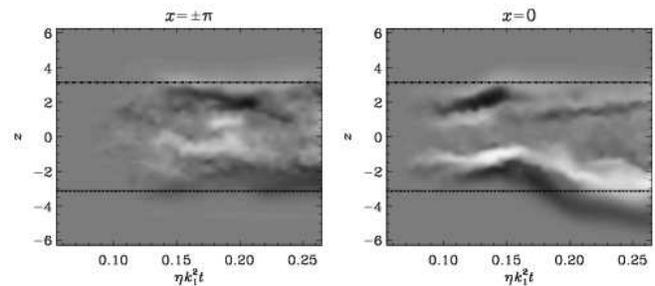}\caption{
Butterfly diagram of the mean toroidal field of \Fig{Fppbym_aft}
at $x=\pm\pi$ and $x=0$.
The lines at $z=\pm\pi$ mark the boundaries of the turbulent slab.
}\label{Fpbutter_ggravs5}\end{figure}

As can be seen in \Fig{Fpbutter_ggravs5}, 
the magnetic field structure is markedly different at $x=\pm\pi$
compared to $x\approx0$, where it is stronger and pushed to the disc
boundaries. This field expulsion to the boundaries could well be the
result of usual field migration that is expected in the presence of shear
and helicity. In terms of mean-field theory with positive $\alpha$ in the
upper disc plane and negative below, we expect field propagation away from
(towards) the midplane when shear is positive (negative). This is
indeed consistent with our model where shear is positive (negative)
at $x=0$ ($x=\pm\pi$).

\section{Conclusions}
\label{Sconcl}

In an attempt to clarify the problem of resistive versus dynamical time scales
in models of the solar dynamo cycle we have tied up a number of loose ends that had been
left open after the first exploratory simulations of B01, BD01, and BBS01.
There are two distinct issues that may or may not be controlled by the
resistive time: the time it takes for the {\it large scale} field to reach
saturation and, once saturation is reached, the time scale on which the
the large scale field can undergo cyclic variations as seen in the sun.

Looking back at \Fig{Fsketch} in the introduction we can still expect various
approaches to be successful, although many others now seem to have been eliminated.
One possibility
was that the dynamo may still operate on a fast time scale if,
like in the kinematic case, the
wavelength of the dynamo wave is shorter than the extent of the system,
and possibly only somewhat larger than the scale of the forcing.
The main problem would then be to keep the
process fast and to prevent the dynamo from developing scales as
large as the box size. How this may happen in reality is not clear.
It may well be connected to the geometry of astrophysical dynamos
which have spherical or disc geometry and are not box-like.
This corresponds to Alternative~A
in \Fig{Fsketch}. The argument against this possibility is that even at
intermediate scales the evolution towards larger scales (i.e.\ the motion
of the secondary bump in the spectrum,
cf.\ \Figs{Fpspec_allo_Hyp3b}{Fpkpeak_comp})
seems to occur on a resistive time scale.
This can be seen from the reduction of the speed
of the spectral bump, $\alpha_{\rm trav}$, as the resistivity
is reduced (\Tab{Tsum}).
So the forcing scale may have to be not much smaller than the
scale of the dynamo wave.
Other plausible mechanisms could be based on non-$\alpha$
effect dynamos. Examples are negative magnetic diffusivity effects
(cf.\ Zheligovsky \ea 2001), the incoherent $\alpha$-effect (Vishniac \&
Brandenburg 1997), and the Vishniac--Cho effect (Vishniac \& Cho 2001). 
These are all subsumed under Alternative~B in \Fig{Fsketch}.

The remaining alternatives are all based on conventional $\alpha$ effect
dynamos, so they have finite net magnetic helicity (although of opposite
sign in the two hemispheres). In order for the dynamo to operate
fast enough, resistive magnetic reconnection (with magnetic helicity destruction)
has to be fast enough. This could only be the case near the surface where the
microscopic magnetic diffusivity is sufficiently large.
This corresponds to Alternative~C in \Fig{Fsketch}.

The other possibility is that the dynamo may work through losses of
magnetic helicity of {\it opposite sign}, possibly at intermediate scales,
for example
near the surface (e.g.\ in coronal mass ejections and active regions) or
across the equator. This is referred to as Alternative~D. The simulations
presented in \Sec{Sremoval} have shown that this is in principle possible.
The problem here is that, so far, none of these mechanisms have been seen
to occur naturally in any of the simulations investigated so far.
It is therefore important to move toward more realistic simulations,
perhaps in spherical geometry, with strong vertical stratification
which could contribute to keeping the large scale field within the
convection zone via turbulent pumping, but allowing fields above a
certain threshold to escape through the surface.

The purpose of this review was to outline the present status of our
understanding of nonlinear large scale dynamos. Things are developing rapidly with new
simulations appearing every month. So far it has mainly been a process
of elimination, because many different ideas have been around, and
constantly more ideas are appearing.
The least explored case is that of
dynamos in realistic spherical geometries.

Although a number of
direct simulations of MHD turbulence in spherical geometry have been
analysed (Gilman 1983, Glatzmaier 1985, Glatzmaier \& Roberts 1995,
Drecker, R\"udiger \& Hollerbach 2000, Ishihara \& Kida 2000) the issue of
fast versus resistively limited growth has not yet been investigated.
This is indeed not an easy task.
In order to distinguish between the two possibilities, one would
need to verify explicitly whether a large scale field is generated
on resistive times, or on a time scale that is independent of and much
shorter than the resistive time scale.
In addition, if the models show cyclic variability (which is desirable
for solar dynamo models), the cycle length needs to be (asymptotically)
independent of the resistive time if the dynamo is fast.
One reason why such numerical experiments are difficult is that only at
relatively large magnetic Reynolds numbers (which require high numerical
resolution) the resistive time scale becomes sufficiently long so that
it can clearly be distinguished from the dynamical time scale. A relevant
dynamical time scale is the ratio $\lambda/(\eta k_1^2)$ which has to
exceed a value of around 20--30 before one can see that resistively
limited growth or cycle periods have occurred (see B01).
At the same time the
resolution must not be too high because otherwise one may not be able
to run the simulations for long enough before any large scale
field has occurred. The
same is true of hyperresistivity which tends to make the large scale
resistive time scale extremely long so that nothing can be said about
resistively limited growth. This explains why the issue of resistively
limited growth is not yet well understood in more realistic geometries.
Nevertheless, it would be worthwhile reanalysing data from recent,
high-resolution dynamo simulations in the light of resistive limitations
on the duration of the saturation phase and, if applicable, the cycle period.

\acknowledgements
Use of the PPARC supported supercomputers in St Andrews and Leicester (UKAFF)
is acknowledged. KS thanks Nordita for hospitality when this work was begun.
This work was supported in part by the Leverhulme Trust (Grant F/125/AL).
We acknowledge the hospitality of the Institute for Theoretical Physics
at the University of California, Santa Barbara, where this work was completed.
This research was supported in part by the National Science Foundation
under Grant No. PHY99-07949

%r e f

\vfill\bigskip\noindent{\it
$ $Id: paper.tex,v 1.129 2002/04/30 17:52:49 brandenb Exp $ $}

\end{document}